\documentclass{aa}  

\usepackage{graphicx}
\usepackage{txfonts}
\usepackage{multicol}
\usepackage{rotating}
\usepackage{float}

\def\HII{H\,{\sc ii}}

\def\arcmin{\hbox{$^\prime$}}
\def\arcsec{\hbox{$^{\prime\prime}$}}
\def\degree{\ensuremath{^\circ}}

\begin{document}

   \title{Kinematics of the \HII~region NGC 7538 from study of the H$\alpha$ line}

   \subtitle{}

   \author{D. Russeil
          \inst{1}
          \and
          H. Plana\inst{2}
          \and
          P. Amram\inst{1}
          \and
          A. Zavagno\inst{1,3}
          \and
          F. Michel\inst{1}
          }

   \institute{Aix-Marseille Univ., CNRS, CNES, LAM, 13388 Marseille, France
        \and
             Laborat\'orio de Astrofísica Te\'orica e Observacional – Departamento de Ci\^encias Exatas – Universidade Estadual de Santa Cruz, Ilh\'eus, BA 45662-900, Brazil 
        \and     
           Institut Universitaire de France, 1 rue Descartes, 75005 Paris, France 
             }


 
  \abstract
   {}
   {Massive stars impact their surrounding initiating star-formation along their photo-dissociation region. Once the \HII~region is formed it is unclear if and how the second generation of stars impacts its aspect and evolution.}
   {We performed high spectral resolution (R $\sim$ 23400) H$\alpha$ Fabry-Perot observations in five fields covering the Galactic \HII~region NGC 7538 and lead profiles multi-gaussian fitting to extract the parameters as peak intensity, width and velocity. We then analyse the kinematics of the ionised gas building kinematic diagrams and second order structure functions for every field.}
   {The observations reveal a general blue-shifted ionised gas flow larger than 11 km s$^{-1}$ in NGC 7538, consistent with previous studies. Profiles originating from features that are dark in H$\alpha$ due to extinction or from 
outside the region show velocity dispersion larger than the one typically found for the Warm Interstellar Medium. The analysis of kinematic diagrams and second-order structure functions reveals non-thermal motions attributed to turbulence and large-scale velocity gradients. In the direction of the \HII~region itself the turbulence seems to be shock-dominated, with a characteristic scale length between $\sim$ 0.72 and 1.46 pc. In this context, we propose that the kinematics of the central part of the region could be explained by the 
superposition of the outflow coming from IRS1 and a wind bow shock formed ahead IRS6.
}
   {}


   \maketitle
%

   \section{Introduction}
   Massive stars have a profound impact on their surrounding through radiative and mechanical feedback from their UV radiation and winds. They can alter the potential of molecular gas to form future generations of stars (negative feedback) by dispersing their parental molecular cloud (\citealt{Walch12}) or, conversely, as illustrated by the Figure 4 in \citet{Deharveng10}, they can promote (positive feedback) and trigger star formation while sweeping up gas into dense shells and by compressing pre-existing local density enhancements such as pillars or globules, as observed in the famous Pillars of Creation by the Hubble (\citealt{Hester96}) and James Webb space telescopes (\citealt{Dewangan24}).

Regardless of the balance between radiative and mechanical feedback, massive stars create \HII~regions (with a bubble-like geometry), featuring various structures (pillars, globules, dense clumps, bright rims, etc.) observed at the interface (the photo-dissociation region, PDR) between the ionized gas and the molecular cloud/layers. These structures testify to this feedback. In particular, from a morphological point of view, \citet{Tremblin13} show that the dense fronts around the ionized gas in the Rosette and Eagle nebulae are systematically asymmetric with low-density gas, where H$\alpha$ emission is present, underlining the fact that they are compressed by the ionized gas.

In terms of kinematics, the expansion of \HII~regions is expected to occur at the speed of sound in the ionised gas, which is about 10 km s$^{-1}$ (\citealt{Verliat22}). 
If the exciting star is located near the edge of its parental molecular cloud or in a non-homogeneous medium, the \HII~region can have its ionizing gas flowing away. This outflow can then follow the Champagne model (\citealt{Tenorio-Tagle79}, \citealt{Arthur06}) or the blister model (\citealt{Hester96}). In parallel, because the exciting star can possess a strong stellar wind, a stellar wind bubble can form in the \HII~region. Such stellar wind bubbles are usually observed thanks to the mid-infrared bow-shock they can form (e.g. in RCW 120, \citealt{Mackey15}). 
However, \cite{Capriotti01} and \cite{Geen21} show that in \HII~regions, unless the ambient density is high, photodissociation dominates the region's dynamics compared to the stellar wind bubble.

In addition, only few studies have examined the impact of the triggered star formation on the development of its parent \HII~region. Young low mass stars commonly produce visible jets, notably observed as Herbig-Haro (HH) objects, which appear as bows or aligned patches of emission. These HH objects are frequently observed in \HII~regions such as for example in the Orion Nebula (\citealt{McCaughrean23}, \citealt{Bally01}, \citealt{Reipurth88}). Thanks to their energetic flows, HH jets are expected to inject energy (e.g. \citealt{Viti03}, \citealt{Decolle05}, \citealt{Verliat22}), causing, maintaining or amplifying turbulence, creating shocks that compress the gas, leading to the formation of new structures (e.g. condensations, filaments and secondary ionisation fronts), and influencing the propagation of the main ionisation front. In particular, the interaction between jets from young massive stars and the ionised gas in the \HII~region can help to sculpt the morphology of the region, by creating cavities. 
This is what \cite{Kraus06}, \cite{Sandell10}, and \cite{Sandell20} propose for the \HII~region NGC 7538, based on a molecular-line study of an outflow driven by a young source located at the edge of the region. 
In this paper, we propose to investigate this aspect through an analysis of the ionised gas. To this end, we focus on the kinematics of the \HII~region NGC 7538 in order to improve our understanding of the feedback processes that shape the region. 

After a general description of the region (Section \ref{objet}) and the observations (Section \ref{data}), a preliminary analysis of the velocity components is presented in Section \ref{genesec}. Section \ref{kinsec} is devoted to the kinematic analysis of NGC 7538, and the results are discussed in Section \ref{discsec}. Conclusions are drawn in Section \ref{conclusec}.

\begin{figure*}[t]
  \begin{center}
   \includegraphics[scale=0.7,angle=0,clip, viewport = 60 209 528 594]{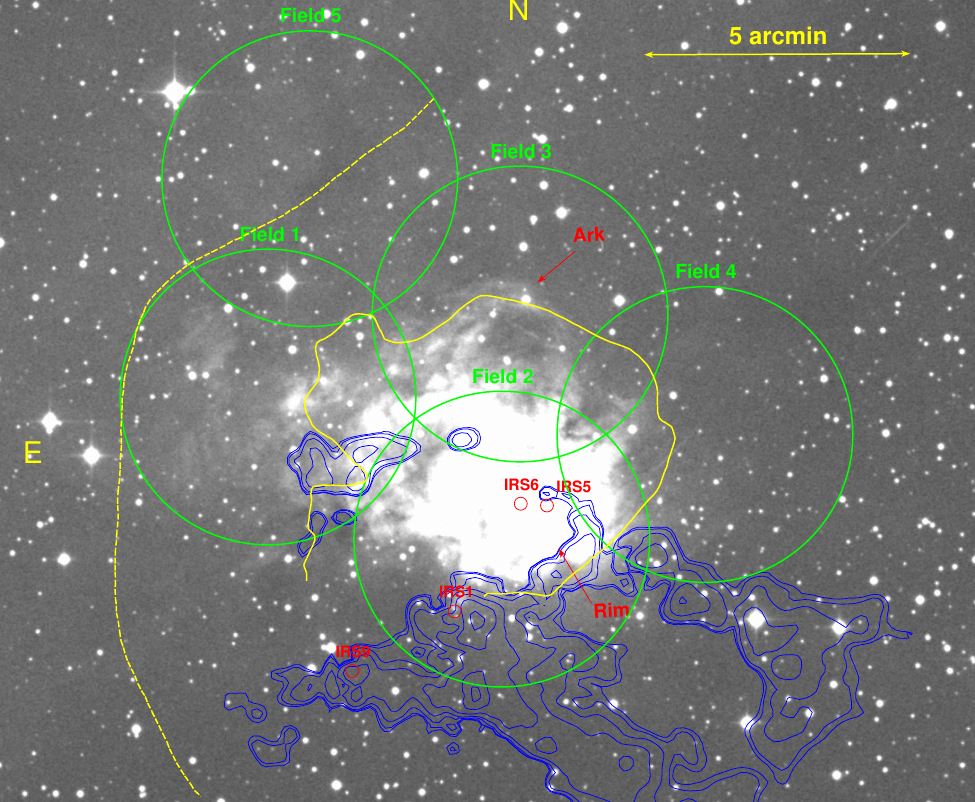}
  \caption{\label{ima1} This image shows the position of the five observed fields overploted on the DSS-Red image of NGC 7538. Yellow isocontours are the inner and the outer (dashed) PDRs boundaries as defined by \citealt{Luisi16} and delineated here from the \textit{HERSCHEL}-70$\mu$m image. The blue isocontours plot the column density (from 1.6 10$^{21}$ to 1.2 10$^{24}$ cm$^{-2}$) constructed from Hi-GAL survey maps (\citealt{Marsh17}). In red are indicated the main infrared sources and the particular features named "Ark" and "Rim". }
   \end{center}
\end{figure*}

   \section{NGC 7538}
   \label{objet}
   The \HII~region NGC 7538 (Sh2-158) has an optical extension of 9\arcmin $\times$ 5\arcmin and is ionized by two binary stars of type O9.5V+B0.5V and O3.5 V((f*)) + O9.5 V (\citealt{Maiz-Apellaniz16}) usually identified as IRS5 and IRS6 respectively (Figure~\ref{ima1}).
   Its associated molecular cloud form a horseshoe surrounding the exciting stars (e.g. \citealt{Barriault07}) but is mainly located at the southwest edge of the \HII~region, as traced by the H$_{2}$ column density distribution (\citealt{Marsh17}), and has a mean LSR velocity of$-$57 km s$^{-1}$ (e.g. \citealt{Dickel81}).
   The \HII~region has shaped a large cavity well identified on Spitzer 8 $\mu$m image, tracing the emission of its surrounding and prominent photodissociation region (PDR).
   \cite{Luisi16} show that there is a well delineated (and closed) main PDR surrounding the \HII~region, while a second arc-like PDR feature is observed to the east, associated to ionising radiation leaking through the main PDR (Figure~\ref{ima1}). In parallel, \cite{Beuther22} show that on the southwest part of the main PDR, the emission is strong and sharp, suggesting compression, and with a layered structure typical of a PDR seen edge-on.

Active star formation is mainly observable to the south, as traced by the infrared sources IRS1 (associated with an outflow; e.g. \citealt{Kraus06}, \citealt{Sandell20}), IRS2, and IRS3 (\citealt{Wynn-Williams74}, associated with UC\HII~regions) and with the infrared cluster NGC 7538S (e.g. \citealt{Bica03}, \citealt{Sandell10}) and numerous clustered YSOs (\citealt{Chavarria14}).
\cite{Sharma17} show that the older ($\sim$ 2 Myr) YSOs population is mainly associated with the \HII~region, whereas the younger ($\sim$1 Myr) one is outside.  
On larger scale \cite{Ma21} show, from CO emission, a clear velocity shear between the east and west parts of the Perseus Arm, divided by a shell-like structure located at l = 111\degree~and \mbox{V$_{LSR}$=-45 km s$^{-1}$} corresponding to the NGC 7538 molecular complex. More precisely, the molecular gas to longitudes larger that the NGC 7538 complex is mainly at$-$37 km s$^{-1}$, while the gas at lower longitudes of the complex is mainly at $-$52 km s$^{-1}$. 
To the NGC 7538 molecular complex can also be associated Sh2-159 and, to the east, an extensive network of filaments and compact sources (\citealt{Reid05}), among which two large elliptical ring structures (\citealt{Fenske21}, \citealt{Fallscheer13}) and the cold cloud G111.80+0.58 (\citealt{Frieswijk07}), all with velocities between $-$56 and $-$46 km s$^{-1}$.    
Finally, the well-known velocity anomaly of the Perseus arm makes the kinematic distance of NGC 7538 invalid and very different from the adopted stellar distance of 2.7$\pm$0.5 kpc (\citealt{Chavarria14}, \citealt{Moscadelli09}, \citealt{Puga10}).

\section{H$\alpha$ Observations and data reduction}
\label{data}

Five regions of NGC 7538 were observed in October 2014 (Figure~\ref{ima1}) using a scanning Fabry-Perot (FP) instrument attached to the 1.93m telescope at the Observatoire de Haute Provence (OHP).
The datacubes were obtained through the GHASP (Gassendi HAlpha Survey of SPirals) instrument \citep{Epinat08, Urrejola22}. This instrument consists of a focal reducer containing a scanning FP interferometer and a GaAs photocathode detector \citep{Gach02}. The field of view is 5.9\arcmin~and the pixel scale is $\sim$ 0.68\arcsec.
The interferometer used is a Queensgate ET70 scanning FP with an order of p=2600@H$\alpha$ enabling a spectral resolution of R $\sim$ 23400. The FP piezos are driven by a CS100 controller, positioned at the telescope.
Observations with the GaAs detector were conducted in interlacing mode, with a 10-second exposure time and 8 cycles of 24 scanning steps. This setup allows us to scan a free spectral range of 2.5 \AA\ (115 km s$^{-1}$) with a scanning step of 0.10~\AA\ (4.8 km s$^{-1}$).
Each field has a total exposure time of 0.53 hours, except for Field 2, which has a 0.47-hour exposure time due to poor weather conditions. The mean seeing during the nights was approximately 3\arcsec.

The data reduction procedure has been extensively described by \citet{Amram96} and \citet{Daigle06a, Daigle06b}.
In this process, a phase map and an instrumental line width (FWHM) map are created from a stable calibration Neon lamp at 6598.95 \AA. The phase calibration map is then used to transform the raw interferograms into wavelength-sorted data at each pixel of the data cubes. The accuracy of the velocity measurements depends on the quality of the phase map. The inspection of the phase map for the five fields reveals no systematic bias and shows only a typical statistical dispersion of 0.7  km s$^{-1}$. In addition, the difference of wavelength between the calibration line and the mean observation (typically of 40 \AA) could introduce a phase shift leading to an other source of uncertainties, with a statistical dispersion of the same order than the previous one, which provides a total statistical dipersion of 1.0 km s$^{-1}$. Once the science data cube has been phase corrected, a light spectral smoothing of 3 channels is performed to improve the SNR. 
All these tasks have been performed using a special package written in IDL\footnote{Package Computeeverything: \url{https://www.astro.umontreal.ca/fantomm/reduction/instructions.html}} \citep{Daigle06b}.
Subsequently, the observed radial velocity is converted into the Local Standard of Rest velocity (LSR). To further improve the SNR, profiles are extracted in 11.7\arcsec $\times$ 11.7\arcsec~(0.15 pc $\times$ 0.15 pc) areas.

The profiles are then decomposed into five components, among which two are the night sky OH and geocoronal H$\alpha$ lines. The three other components, coming from the nebula and the line of sight diffuse emissions, are modeled by Gaussians whose parameters (intensity, position, and FWHM) are fitted using the Minuit\footnote{ \url{https://root.cern.ch/download/minuit.pdf}} minimization tool (\citealt{James94}). The night–sky lines vary significantly with observing conditions (e.g. \citealt{Zhang21}). However, observations typically show a geocoronal H$\alpha$–to–OH line intensity ratio of approximately 2 (\citealt{Haffner03}), while the absolute intensity of the geocoronal H$\alpha$ line generally falls within the range $0.5$--$5 \times 10^{-16} \mathrm{erg s^{-1} cm^{-2} \AA^{-1}}$ (\citealt{Zhang21}).
To optimise these parameters and determine the number of nebular Gaussian components required, we first extracted and manually fitted both the field–averaged spectrum and spectra from low–emission regions where the geocoronal line is of similar intensity to the nebular emission. This procedure enabled us to estimate the widths, intensities, and central wavelengths of the night-sky lines, as well as the minimum number of nebular Gaussian components needed for a reliable fit.
Then, for automated fitting, we constrained the positions of the night-sky lines to a tolerance of 3 km s$^{-1}$, adopted a fixed width, and set the H$\alpha$-to-OH line intensity ratio to 1.7. We also restricted the geocoronal H$\alpha$ intensity to values between 0.1 and 1 (arbitrary units). The fitting configuration enables the algorithm to begin with a single nebular Gaussian, with the other components initially set to zero. However, as with any decomposition, there is an inherent uncertainty when decomposing such broad and blended profiles. In addition, because the filter transmission {\bf decreases to about 40\% from the center of the field to its edge,} profile fitting becomes more uncertain in the field’s outer regions and particularly in areas of low emission. 

\begin{figure}[h]
  \includegraphics[scale=0.42,angle=0,clip, viewport = 70 4 639 435]{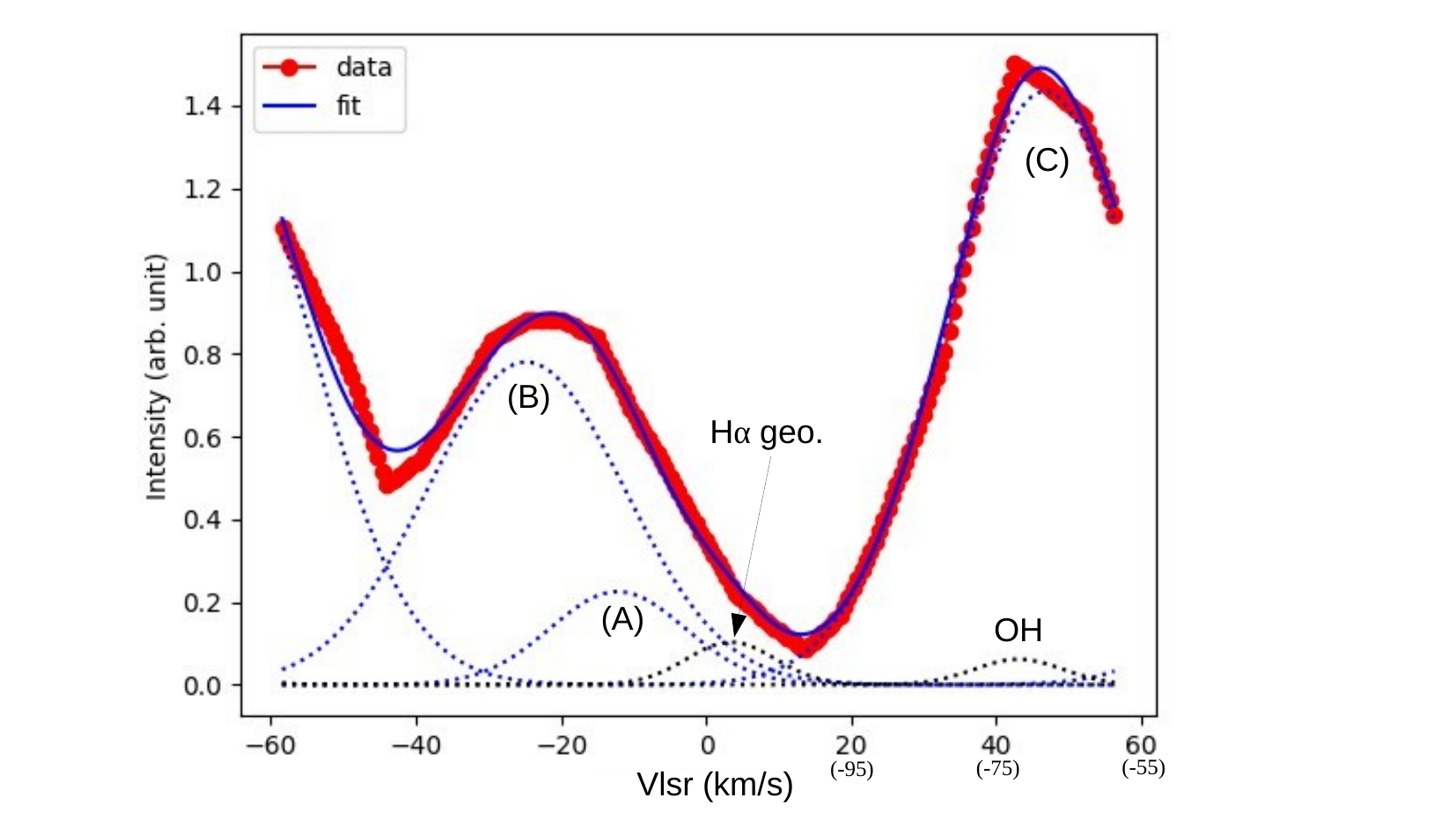}
  \caption{\label{prof} Example of profile decomposition. The blue and black dotted lines are the H$\alpha$ emissions (labeled A, B and C) and  night-sky (labeled "H$\alpha$ geo." and "OH") components respectively. On the x-axis the velocities corrected for the free spectral range are shown in parentheses. The blue and the red plots are the observed
    and the fitted profiles, respectively.}
\end{figure}

An example is shown in Figure \ref{prof}. In the following, the uncertainties on the Gaussian parameters are evaluated following \citet{Lenz92}, and the FWHMs (hence the velocity dispersion $\sigma$) are deconvolved from the instrumental widening using the instrumental line width map.
Note that the data are not flux-calibrated; therefore, the intensity is in arbitrary units (a.u.). Note that the instrumental configuration causes the OH lines at 6554 \AA~and 6569 \AA~to be superimposed on the same spectral/velocity channel and that velocities are determined modulo the free spectral range. This implies that  throughout the rest of the paper the positive velocities (e.g. on the right side of the Figure \ref{prof}) are in fact negative (by subtraction of the free spectral range).

\section{General overview of the velocity components}
\label{genesec}

\begin{figure}[]
  \includegraphics[scale=0.32,angle=-90,clip, viewport = 45 48 567 781]{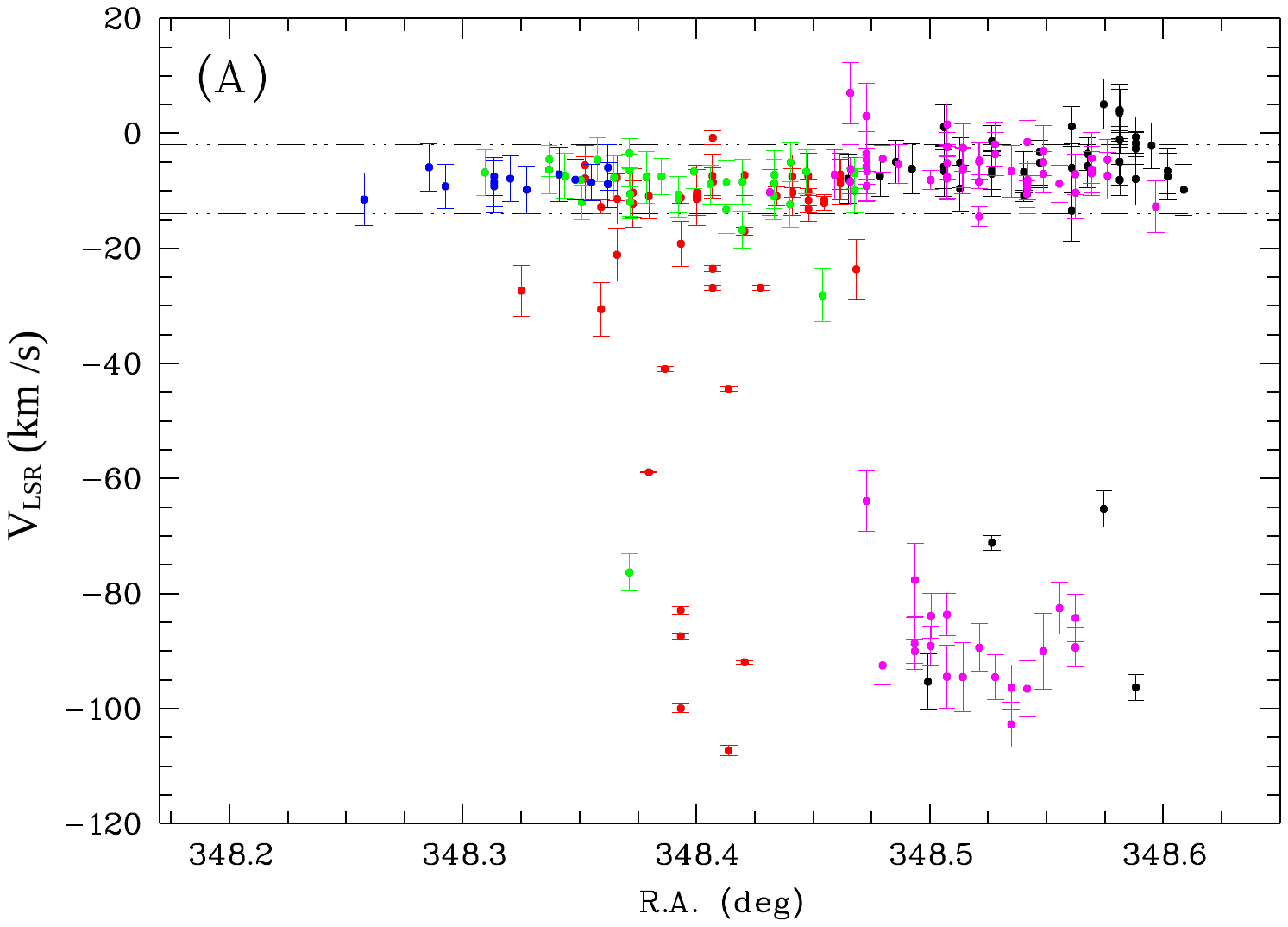}
  \includegraphics[scale=0.32,angle=-90,clip, viewport = 45 48 567 781]{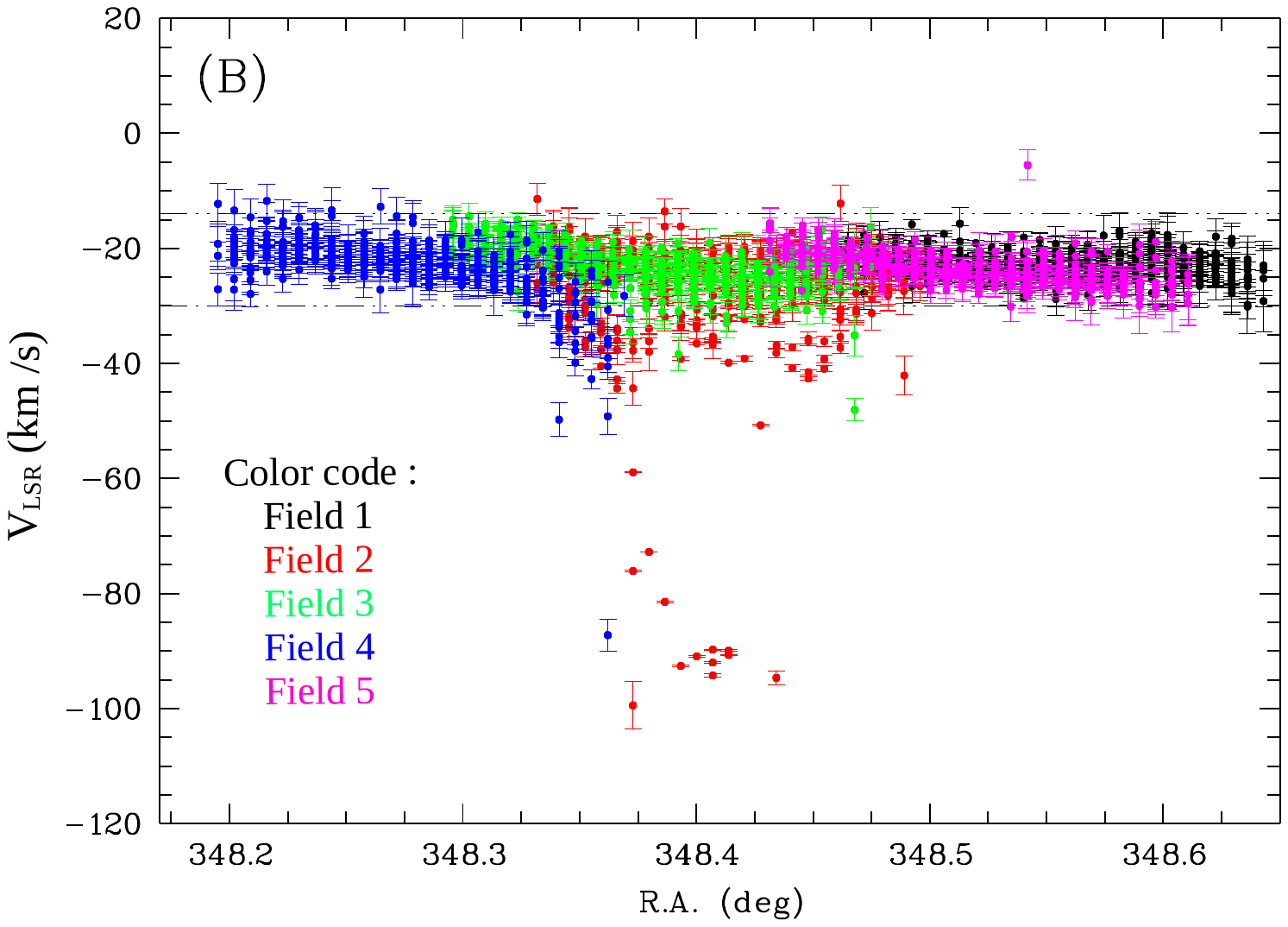}
  \includegraphics[scale=0.32,angle=-90,clip, viewport = 45 48 567 781]{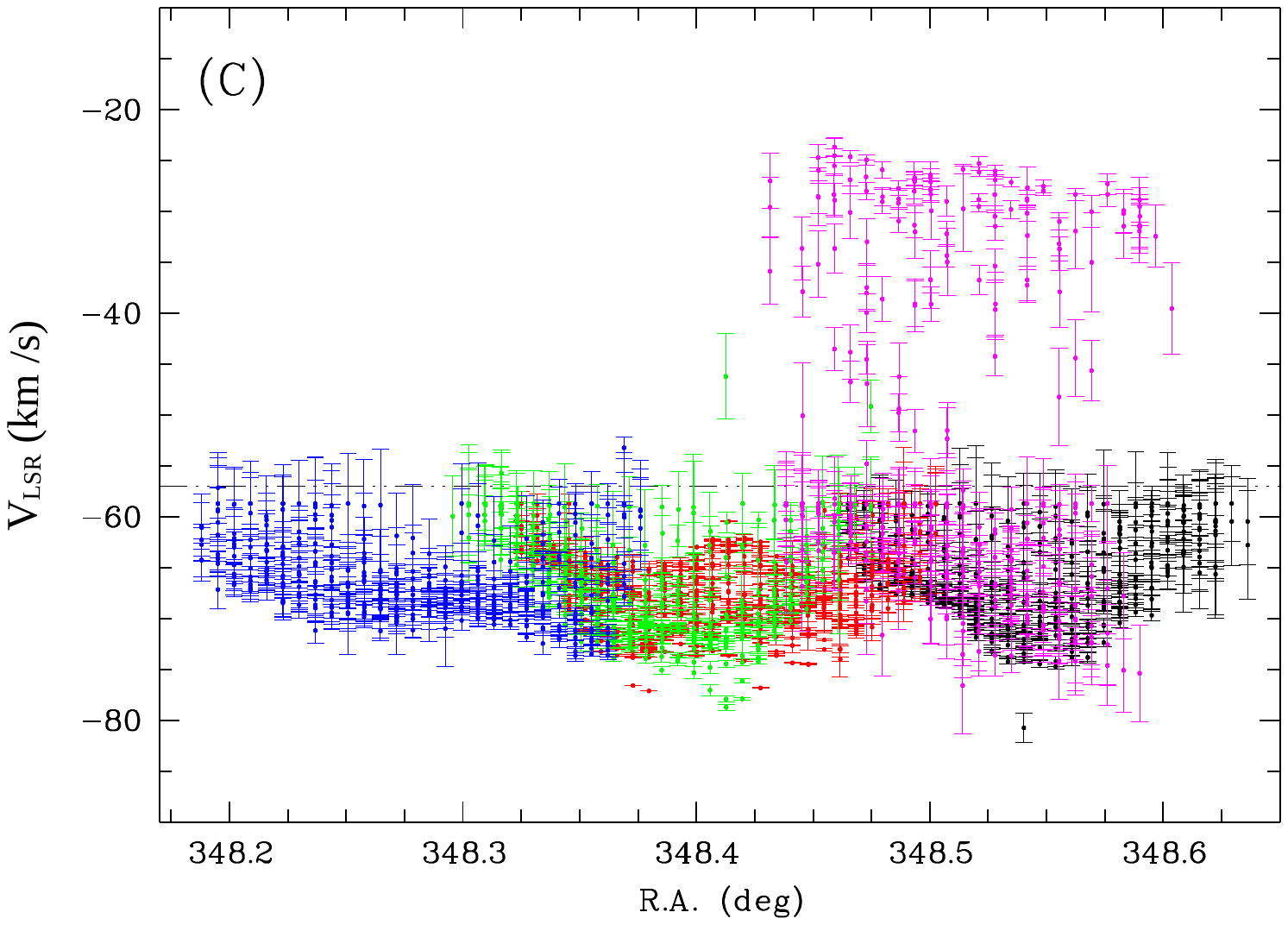}  
  \caption{\label{diag1} Position-velocity plots of the three fitted Gaussians arranged in ascending order of intensity (from panel (A) to (C)). The black, red, green, blue and magenta correspond to the data from the fields 1 to 5, respectively.}
\end{figure}

Once the fitting is performed, the three different nebular components are sorted by decreasing peak intensity. Figure \ref{diag1} presents the distribution of each component in a V$_{LSR}$ versus right ascension plot. We have retained only the significant and reliable data: peak intensities greater than 0.2 and $\sigma$ between 6 and 35 km s$^{-1}$. 

Figure \ref{diag1} (A) to (C) shows the position-velocity plots of the fited components A, B and C, respectively, and help us to distinguish the different layers present along the line of sight. First, a barely detected emission with a mean V$_{LSR}$ of $-$7.2~$\pm$~3.8 km s$^{-1}$ (Figure \ref{diag1}(A)) is observed and is a local component. On Figure \ref{diag1}(B) we clearly identify (three times more intense than the local one), centered at $-$23 $\pm$ 3.1 km s$^{-1}$, a foreground warm interstellar medium emission. In Figures \ref{diag1}(B) and \ref{diag1}(C), a small fraction (3\%) of measurements are around $-$36.9 km s$^{-1}$ (and 1.6 times more intense than the $-$23 km s$^{-1}$ emission), and these originate mainly from Fields 2 and 4. In the following, we call that emission the "$-$36 km s$^{-1}$" component. On Figure \ref{diag1}(A) and Figure \ref{diag1}(B), a few measurements (0.3\%) are found around $-$92 km~s$^{-1}$; they are localized at the \HII~region center. In the following, we call that emission the "$-$92 km~s$^{-1}$" component. Finally, from Figure \ref{diag1}(C) with velocities between $\sim$$-$57 km~s$^{-1}$ and $\sim$$-$80 km~s$^{-1}$ we identify the "main" velocity component from NGC~7538 while in Field 5 (magenta), the $-$23 km~s$^{-1}$ emission appears to be the most intense.
To place these velocity components in the context of the spiral arms, we refer to the $^{12}$CO emission (from 105\degr~< l < 119\degr, and $-$5\degr~< b < 5\degr) observed by \cite{Ma21}. These data delineate the spiral arms as follows: the Local Arm (from $-$27 to +20 km s$^{-1}$), the Perseus Arm (from $-$75 to $-$27 km s$^{-1}$), and the Outer Arm (from $-$115 to $-$75 km s$^{-1}$). In parallel, \cite{rigby24} shows that the bulk CO emission from NGC~7538 is contained within $-$70 to $-$40 km s$^{-1}$, with a secondary minor emission component between $-$17 and $-$3 km s$^{-1}$ is also detected along the line of sight.

Additionally, based on molecular clouds within 40\arcmin, around NGC 7538 listed by \cite{Miville17} and \cite{Brunt03}, we can note three distinct velocity groups: clouds with velocities between $-$5 and $-$16 km s$^{-1}$, which are typically associated with dark features in optical images; a few clouds with velocities between $-$29 and $-$35 km s$^{-1}$, located mainly south of the \HII~region; and another group with velocities between $-$42 and $-$68 km s$^{-1}$. From the WHAM survey (\citealt{Haffner03}), two velocity layers of H$\alpha$ emission are detected between Galactic longitudes of 100\degr~to 120\degr: a local layer around $-$11 km s$^{-1}$ and another around $-$41 km s$^{-1}$, primarily located at b $\sim$ $-$0.78\degr.

In this context, we can attribute the H$\alpha$ emissions at $-$7.2 km s$^{-1}$ and $-$23 km s$^{-1}$ to foreground warm interstellar medium. Furthermore, to explain the systematic presence of the $-$23 km s$^{-1}$ component and its lack of a clear associated molecular emission, we propose that it could trace the warm ionized medium inside the large (2\degr) HI shell GS112+1-020, identified by \cite{Ehlerova05} at l, b = 111.7\degr, +1.3\degr, and with a velocity of $-$19.5 km s$^{-1}$ (and a velocity range of 12.4 km s$^{-1}$).

Since these two foreground emissions lie outside the scope of this study, they are excluded from the subsequent analysis while the "main" (velocities between $-$57 km s$^{-1}$ and $-$80 km s$^{-1}$), the "$-$36 km s$^{-1}$" and the "$-$92 km s$^{-1}$" components are thought to originate from NGC~7538.

\begin{figure*}[]
\includegraphics[scale=0.14,clip, viewport = 1 1 1290 811]{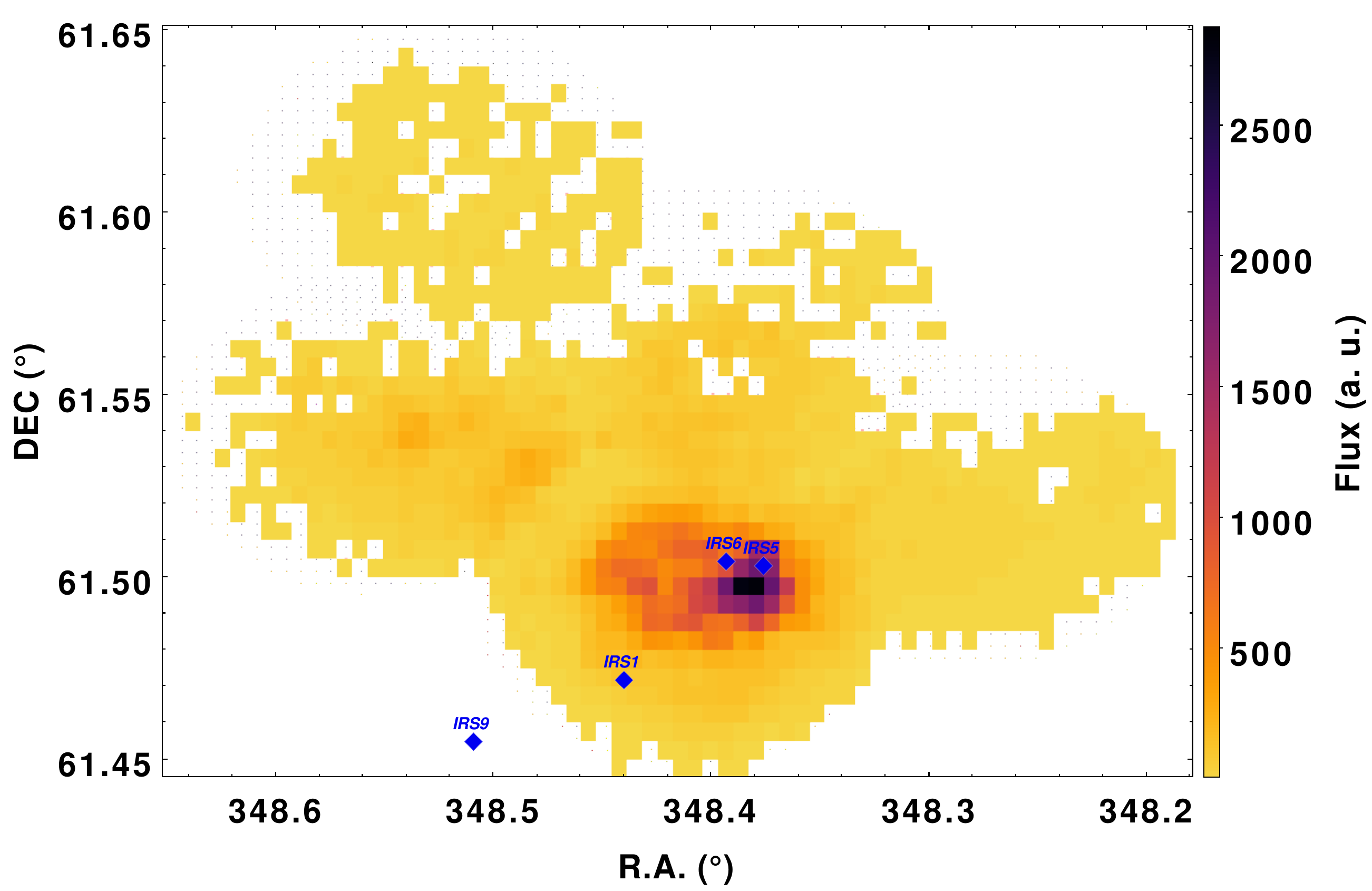}\includegraphics[scale=0.14, clip, viewport = 49 1 1290 811]{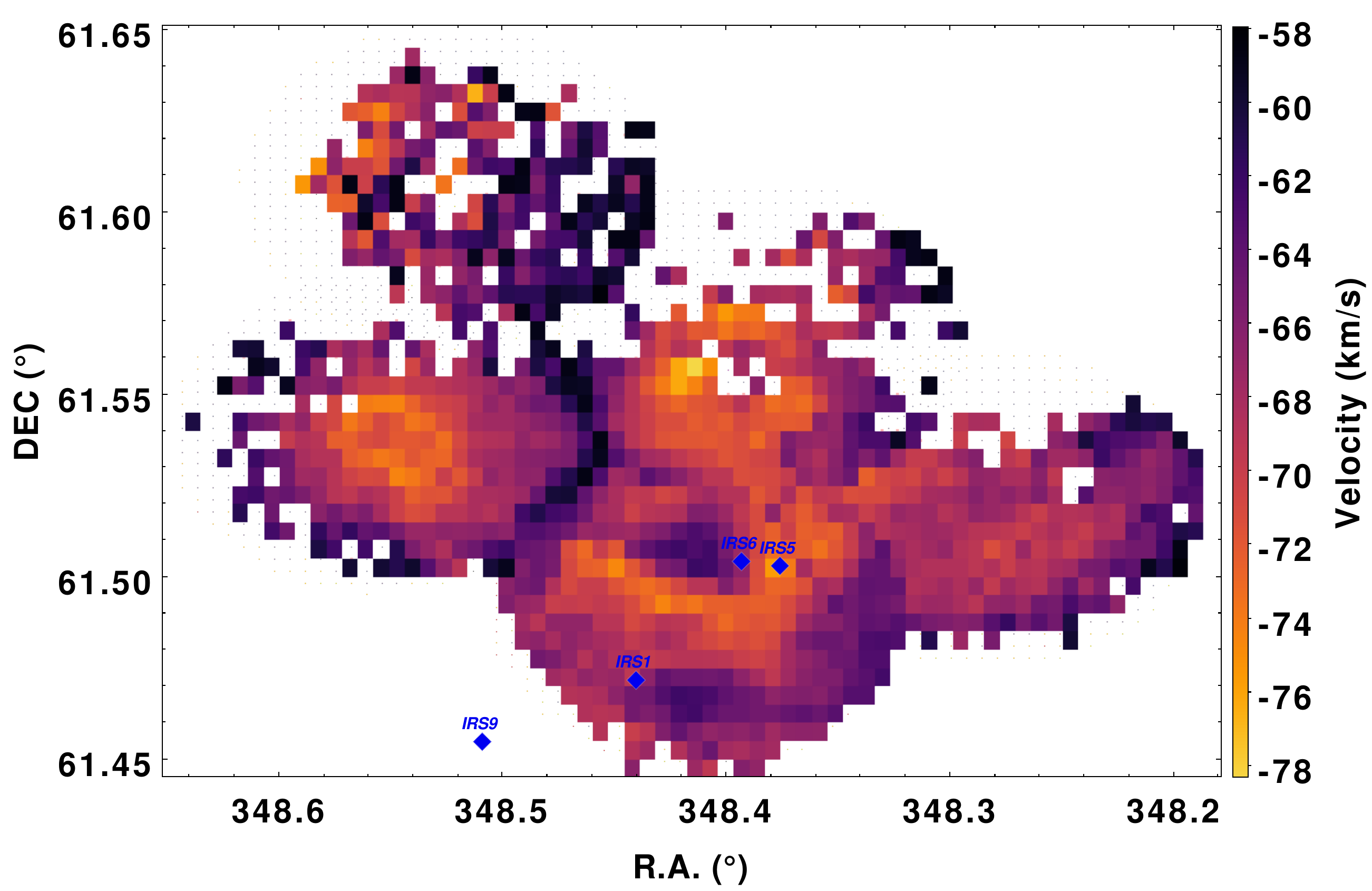}\includegraphics[scale=0.14, clip, viewport = 49 1 1250 811]{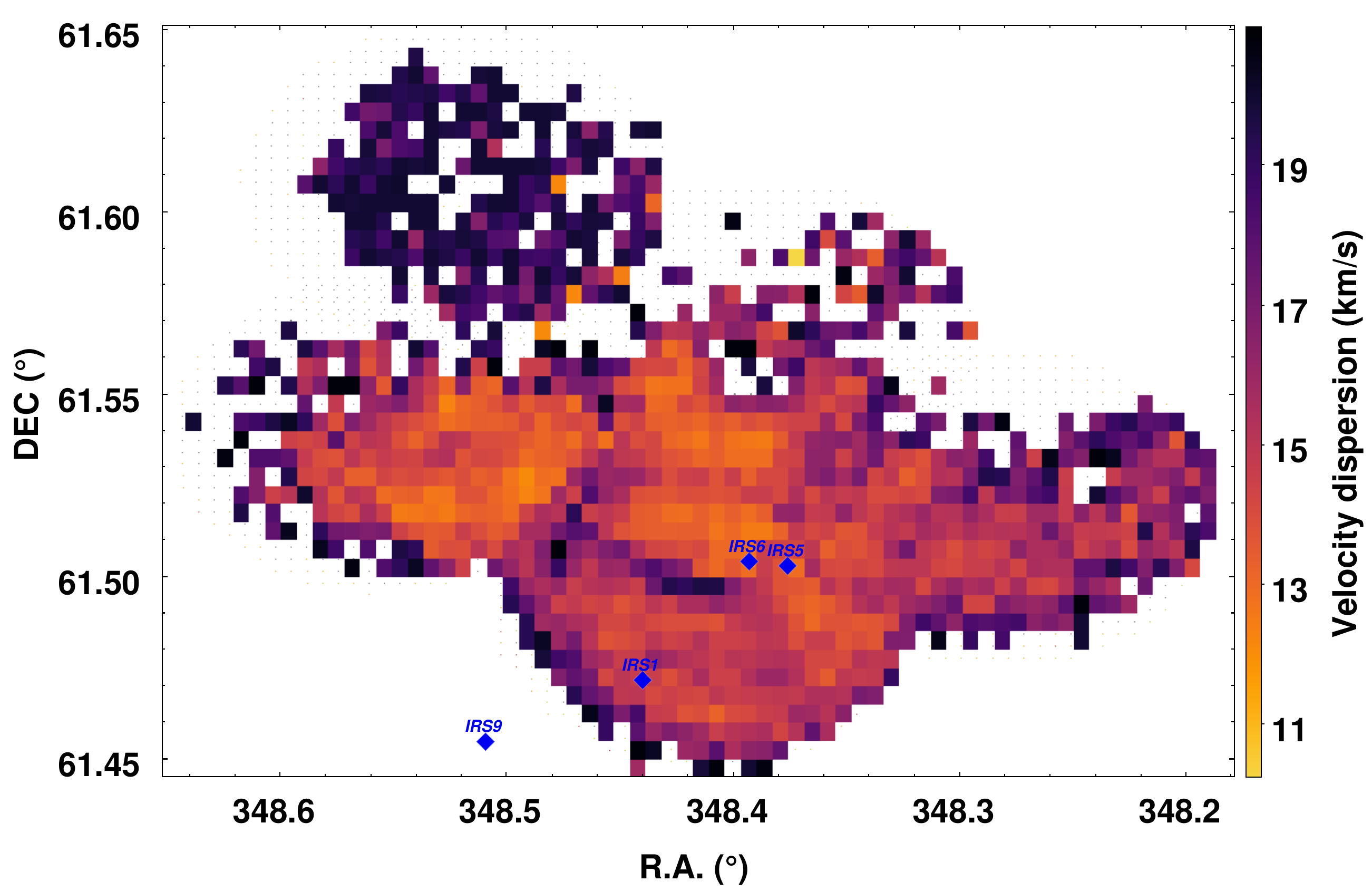}
\includegraphics[scale=0.14,clip, viewport = 1 1 1290 811]{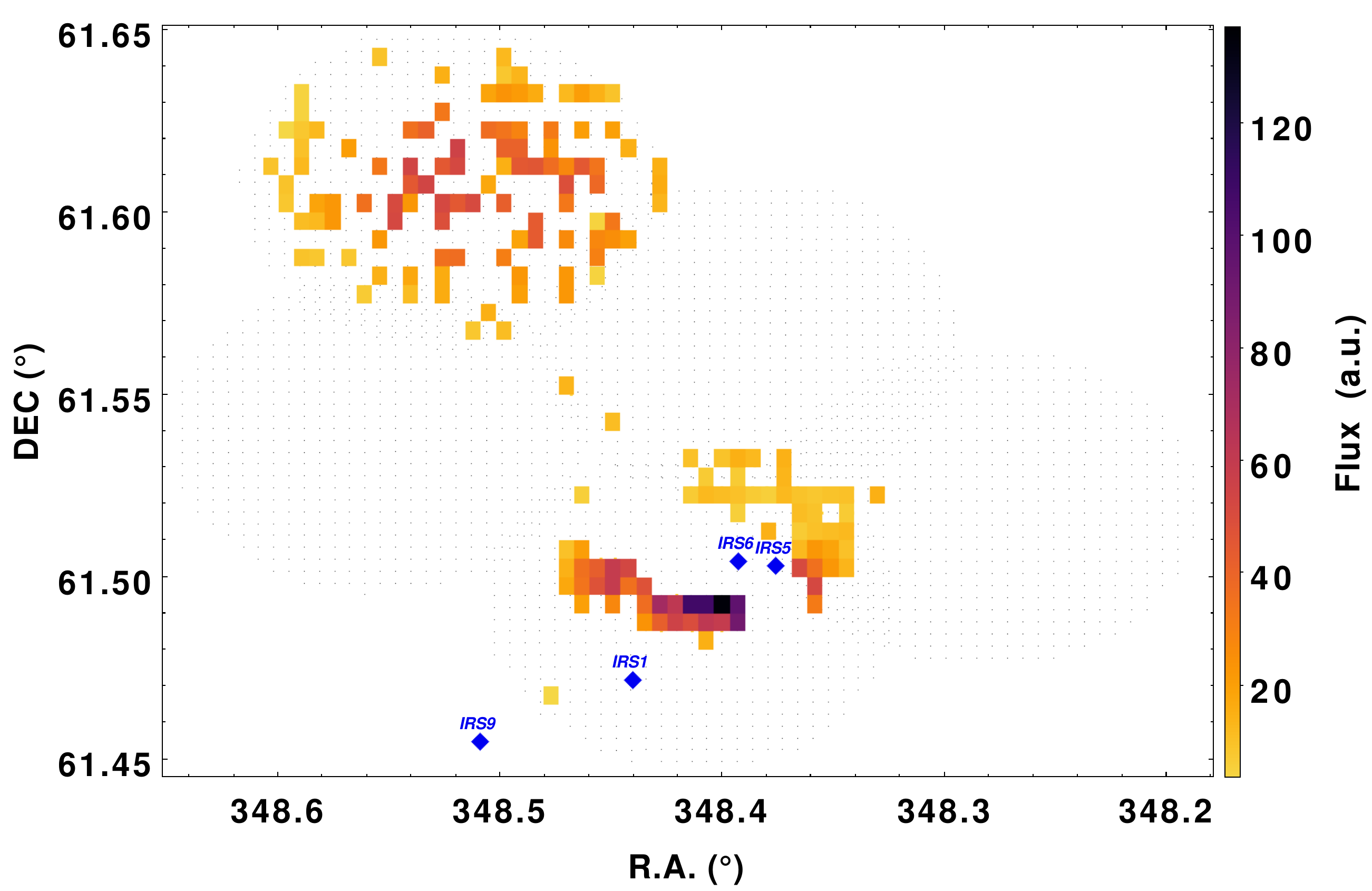}\includegraphics[scale=0.14, clip, viewport = 49 1 1290 811]{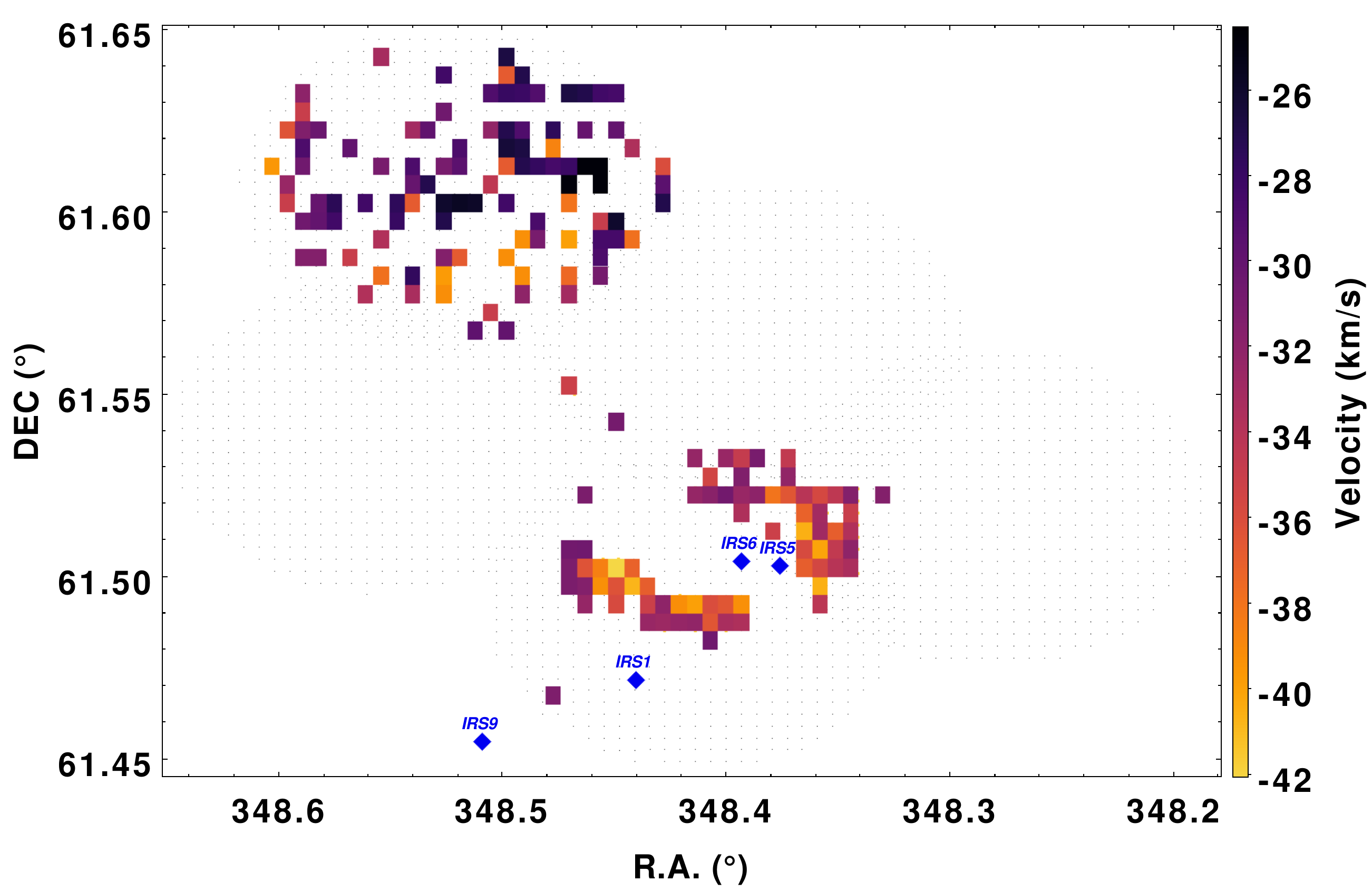}\includegraphics[scale=0.14, clip, viewport = 49 1 1250 811]{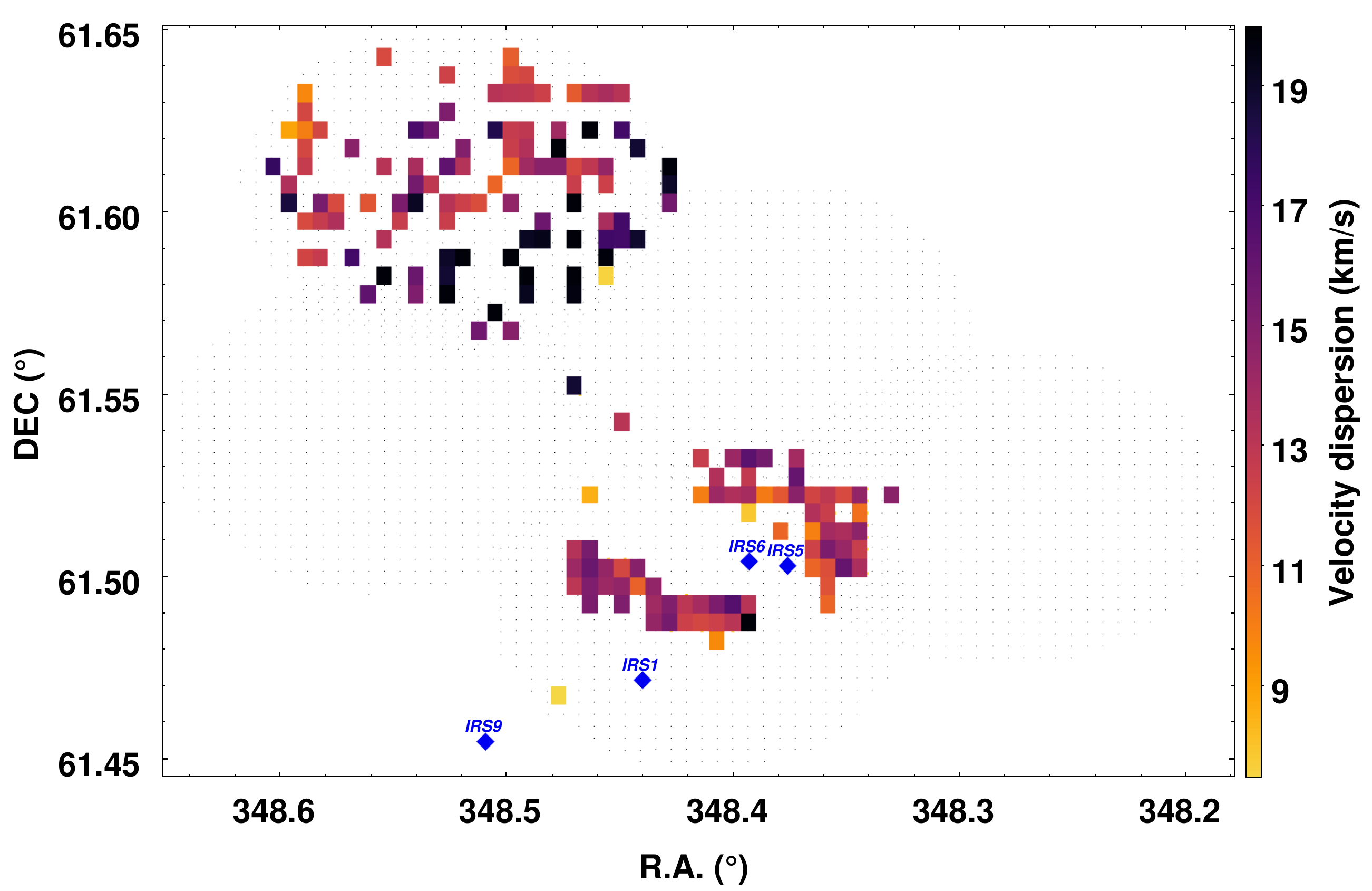}
\includegraphics[scale=0.14,clip, viewport = 1 1 1290 811]{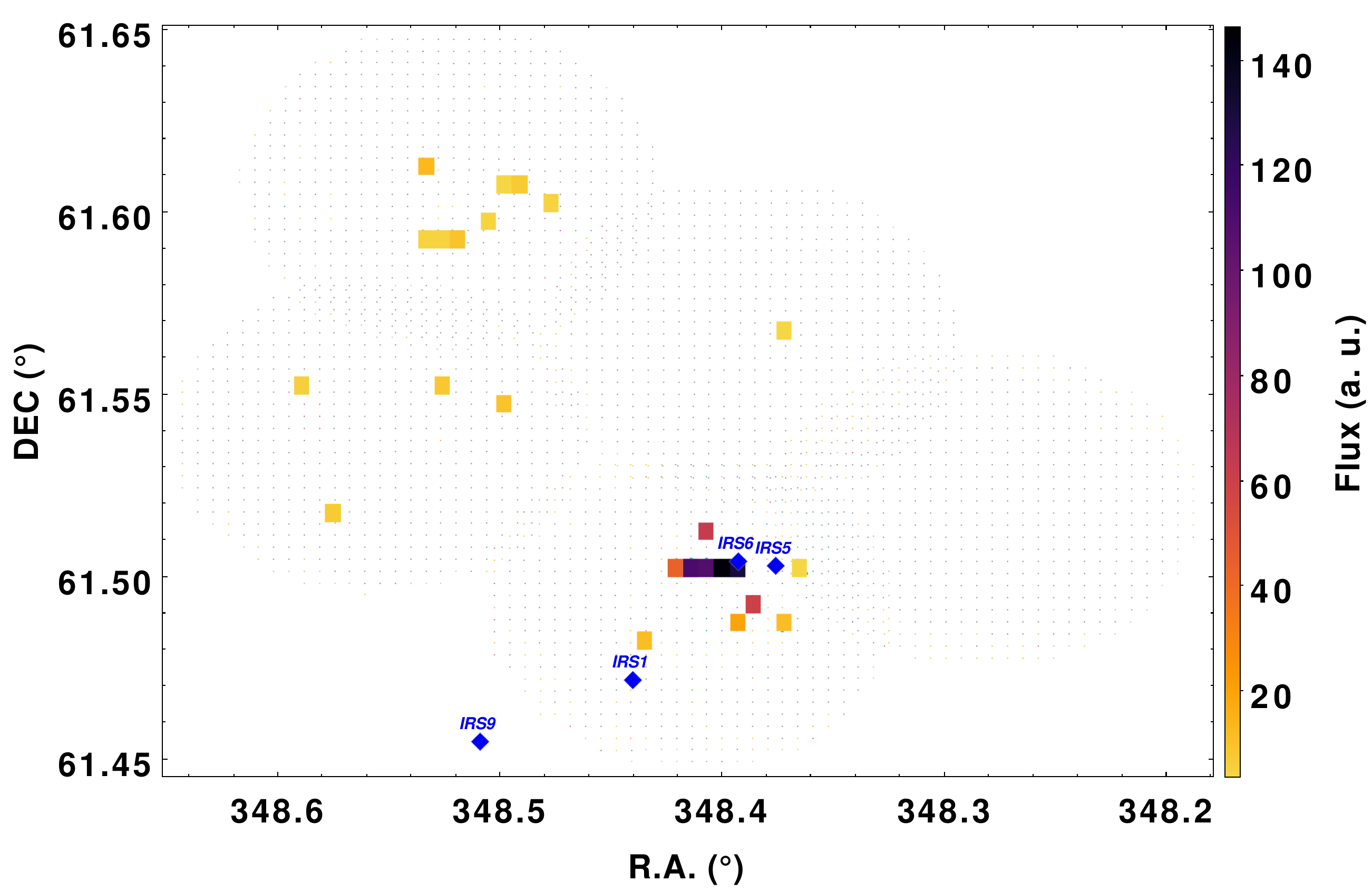}\includegraphics[scale=0.14, clip, viewport = 49 1 1290 811]{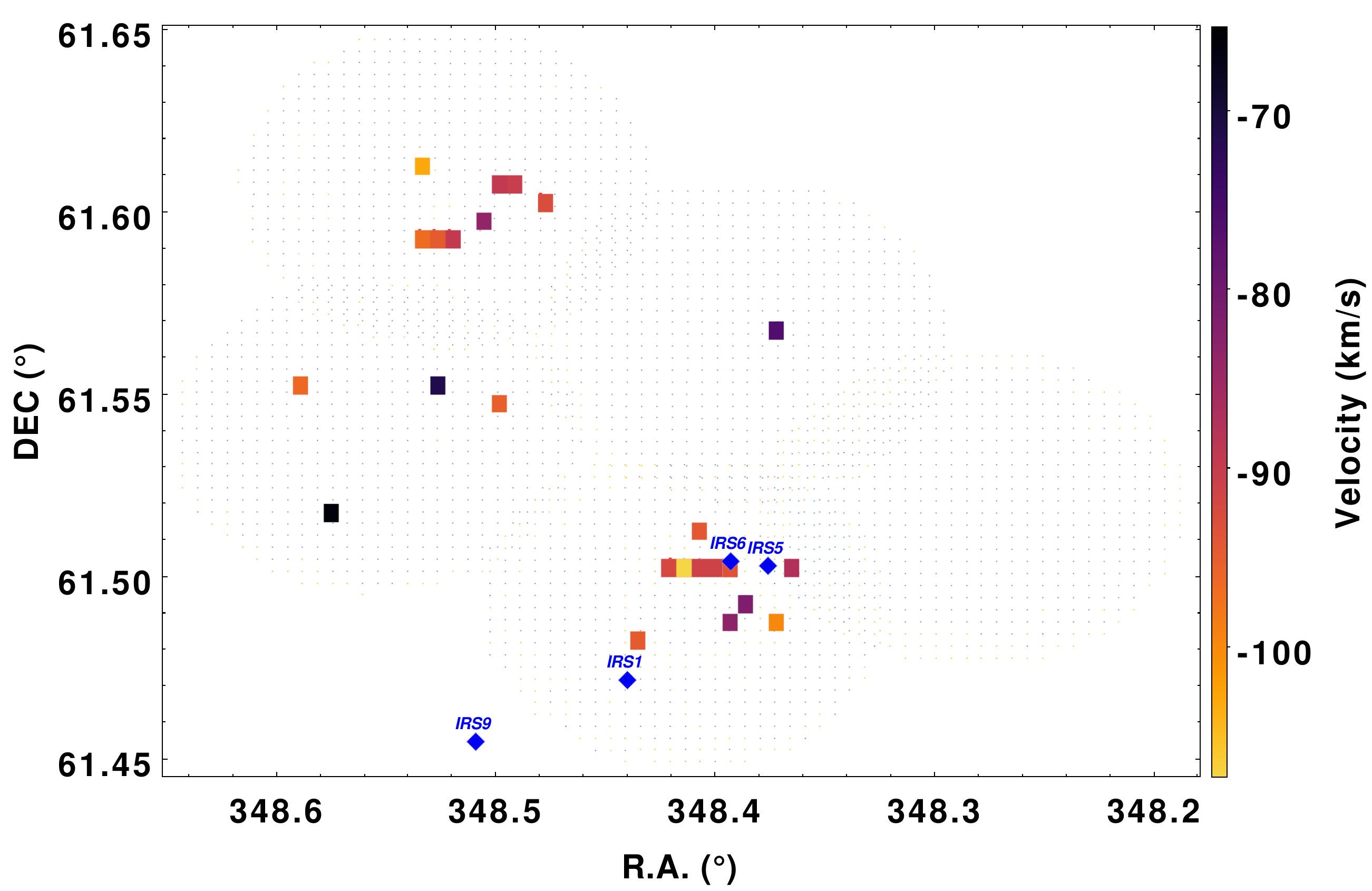}\includegraphics[scale=0.14, clip, viewport = 49 1 1250 811]{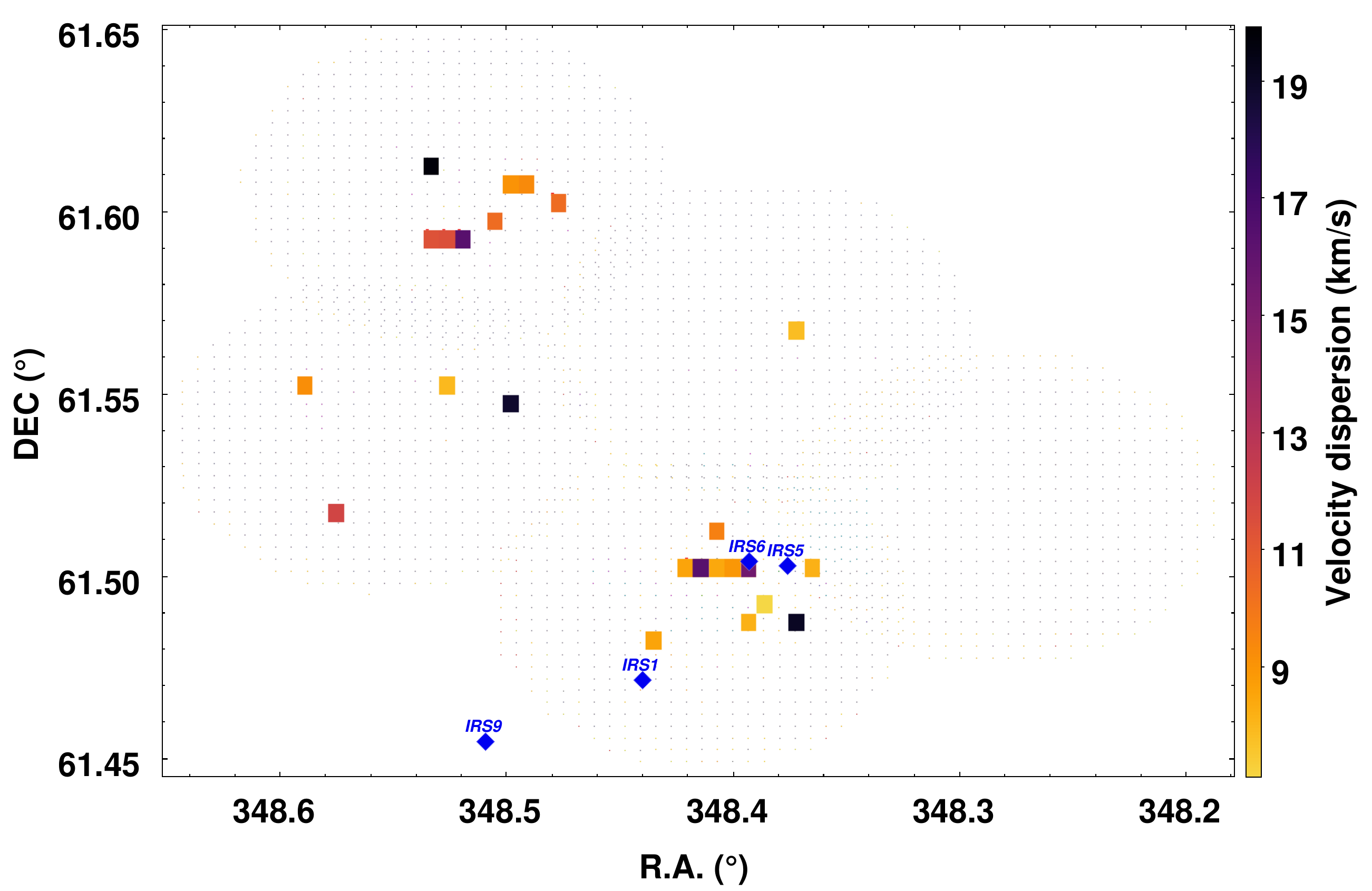}
\caption{\label{maps} Flux (left), velocity (middle) and velocity dispersion (right) maps for the main component (top line panels), the "$-$36 km s$^{-1}$" (middle line panels) and the "$-$92 km s$^{-1}$" (bottom line panels) components respectively. The values in the overlapping areas (particularly in the top panels) between the different fields are simply an average. The position of IRS1, IRS5, IRS6 and IRS9 are indicated as blue symbols.}
\end{figure*}

Figure \ref{maps} provides an overview of the fitted parameter maps corresponding to the velocity components of NGC~7538 with peak intensity greater than 0.15 (arbitrary units) to retain the most intense ones (typically profiles with signal-to-noise ratio~>~3) and keeping the ones with realistic width (between 8 km~s$^{-1}$ and 30 km s$^{-1}$). Indeed as noted in Section~\ref{data}, part of velocity discrepancies observed at the field overlaps may result from uncertainties associated with fitting low-emission profiles. To mitigate these effects, we primarily adopted a field-by-field approach systematically selecting the most reliable components.

\section{NGC 7538 ionised gas kinematic}
\label{kinsec}

In this section, we focus on the main component (velocities between $-$57 km s$^{-1}$ and $-$80 km s$^{-1}$).

\subsection{Kinematic diagrams analysis}
\label{diag}

\begin{figure*}[]
\includegraphics[scale=0.243,angle=-90,clip, viewport = 46 71 575 788]{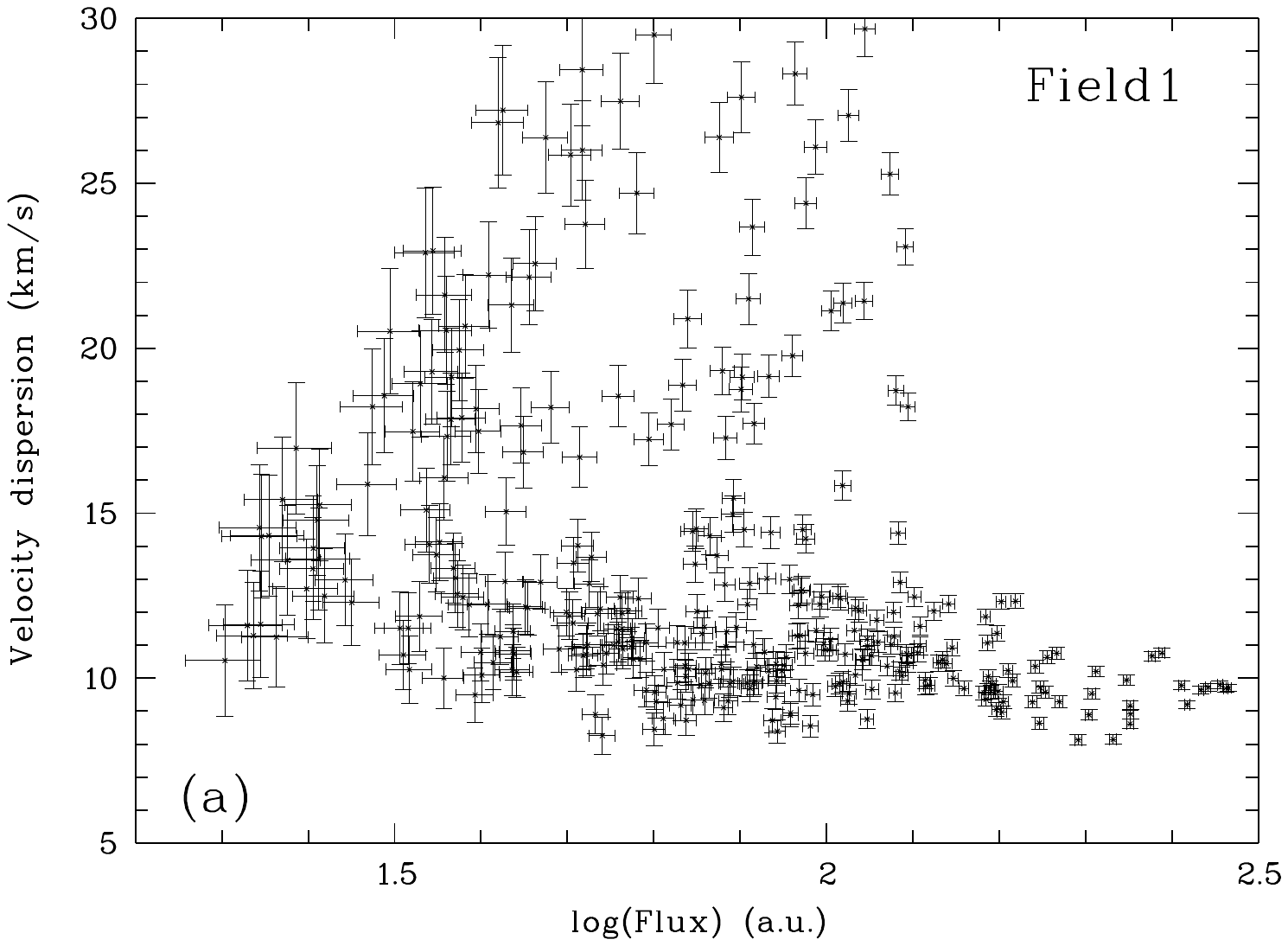} \includegraphics[scale=0.243,angle=-90,clip, viewport = 46 60 575 788]{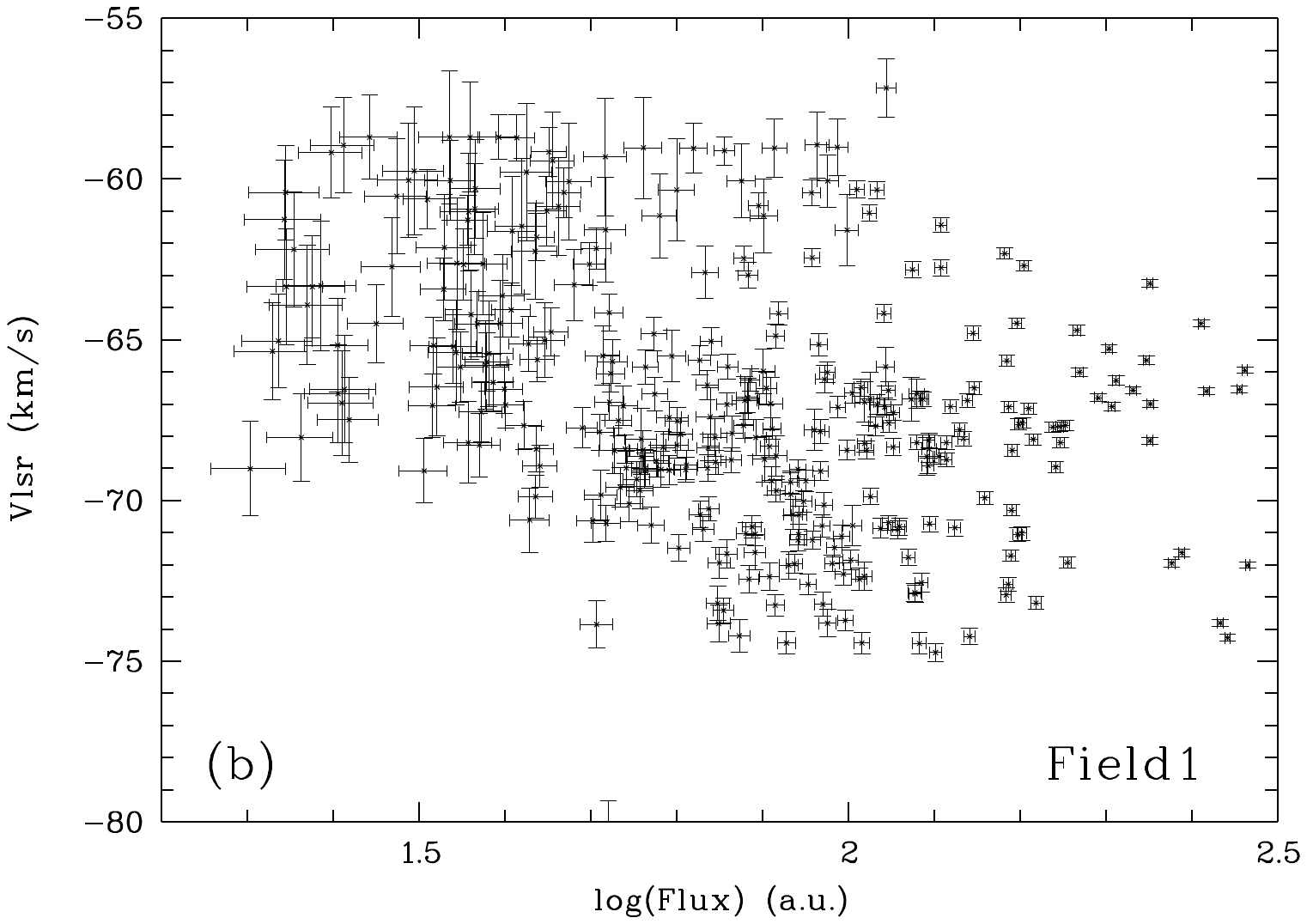} \includegraphics[scale=0.243,angle=-90,clip, viewport = 46 71 575 788]{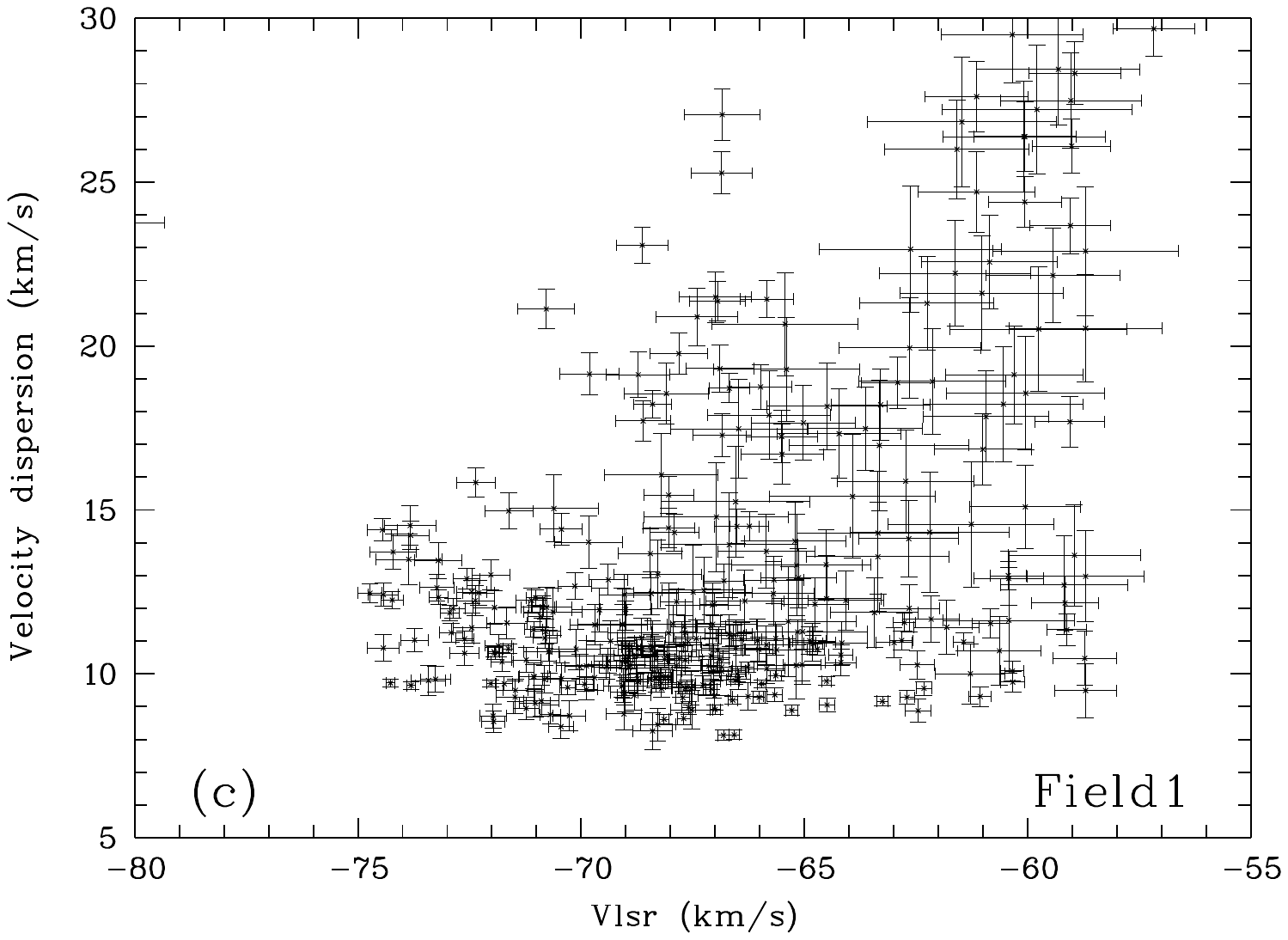}
\includegraphics[scale=0.243,angle=-90,clip, viewport = 46 71 575 788]{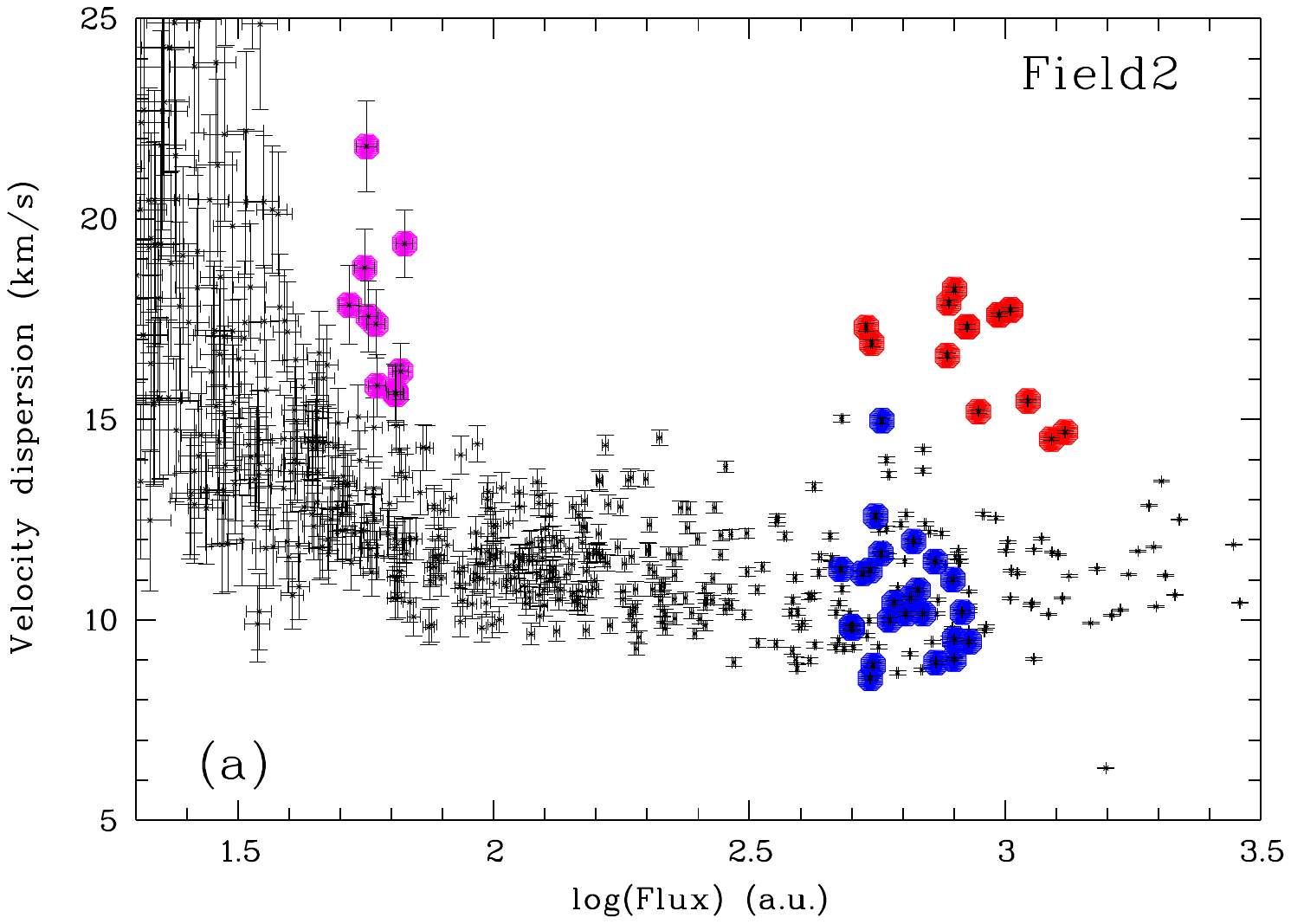} \includegraphics[scale=0.243,angle=-90,clip, viewport = 46 60 575 788]{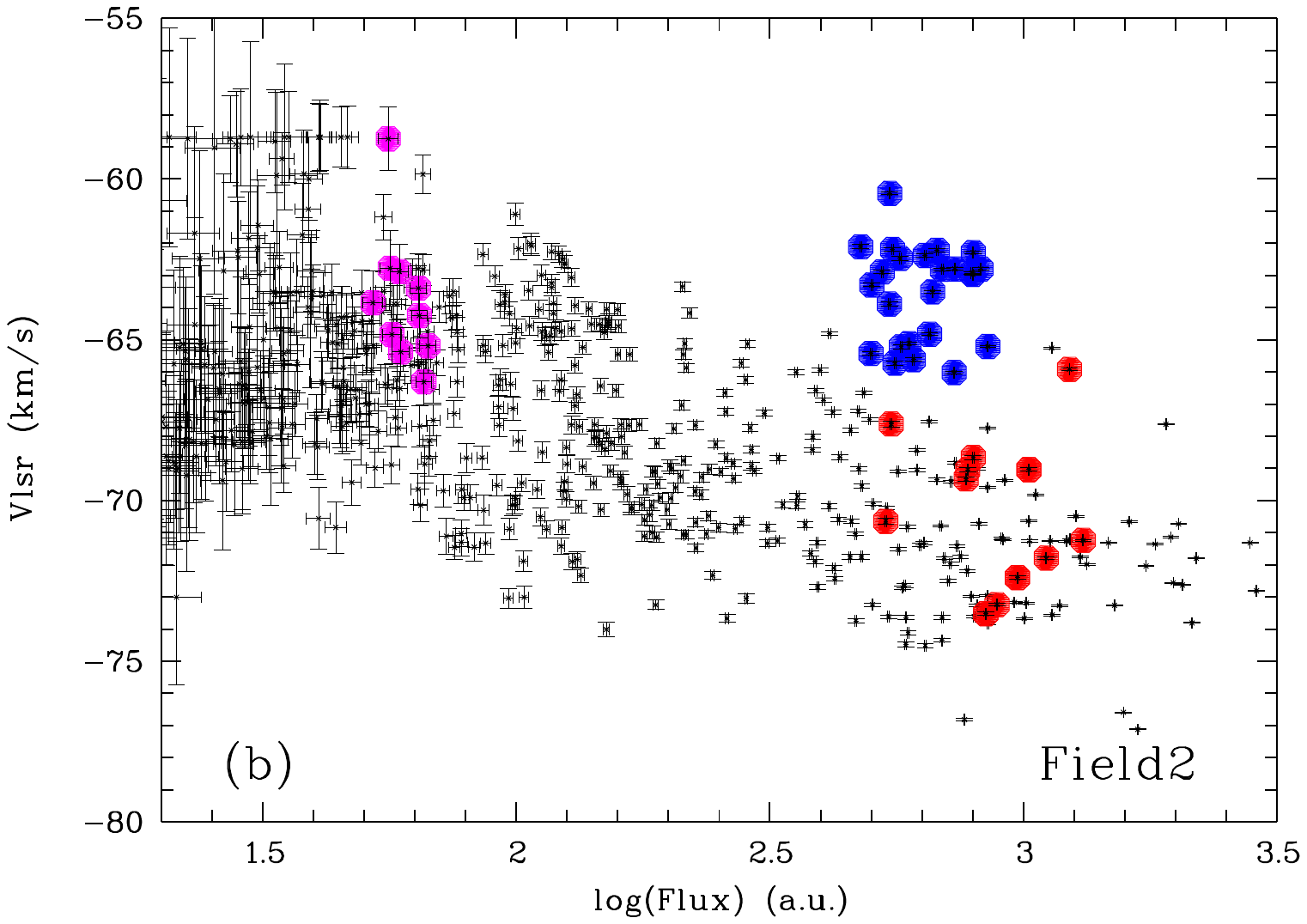} \includegraphics[scale=0.243,angle=-90,clip, viewport = 46 71 575 788]{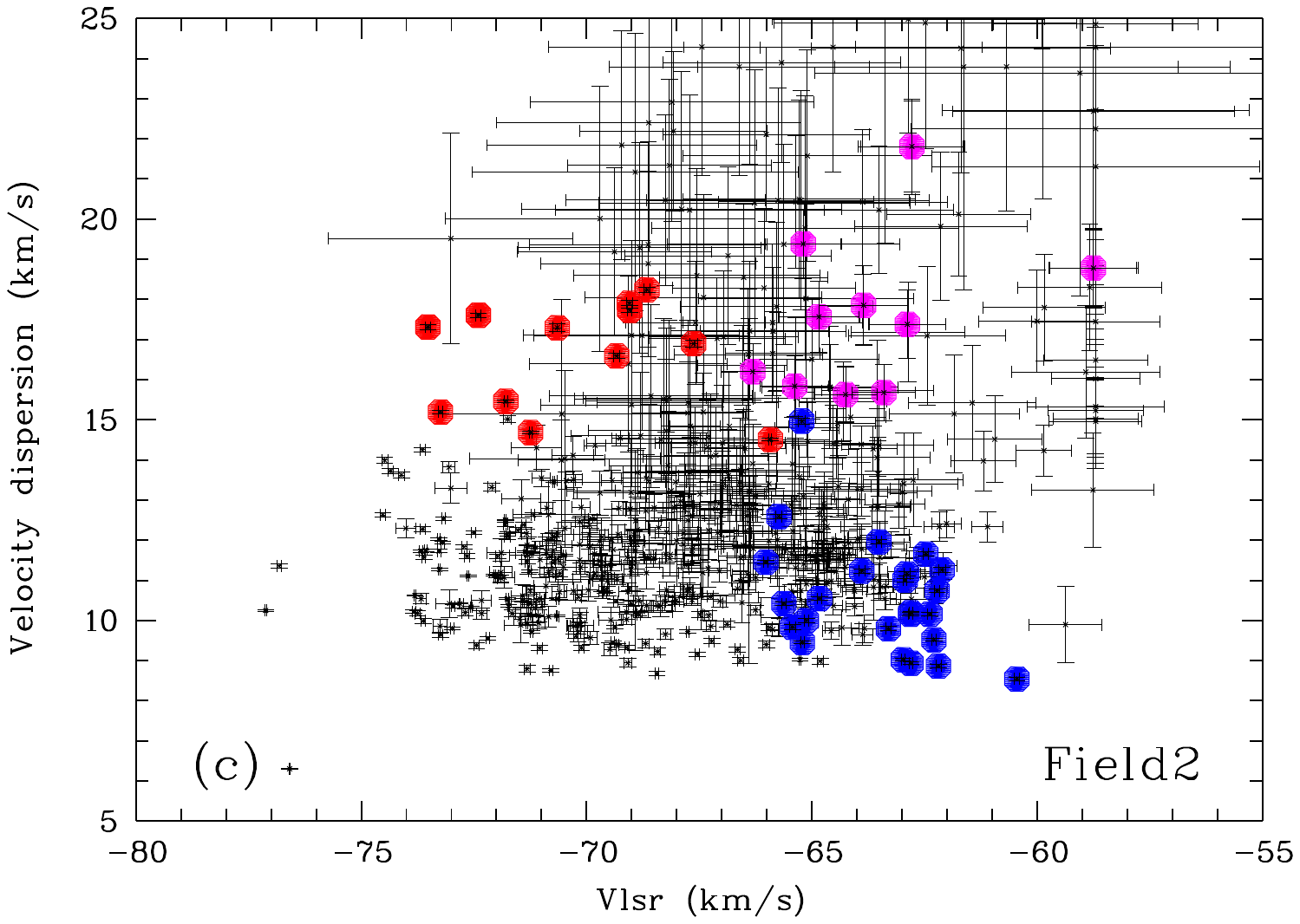}
\includegraphics[scale=0.243,angle=-90,clip, viewport = 46 71 575 788]{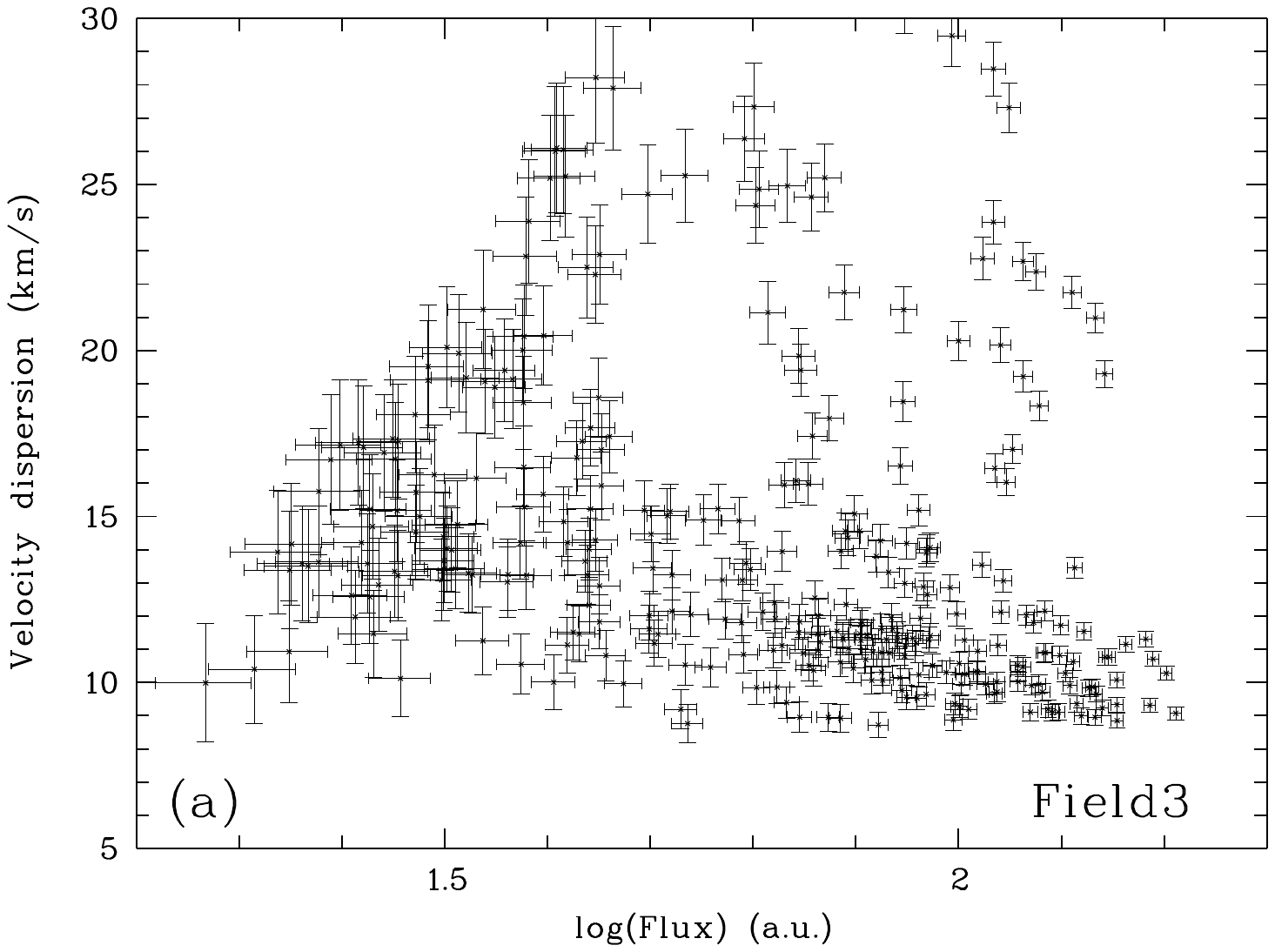} \includegraphics[scale=0.243,angle=-90,clip, viewport = 46 60 575 788]{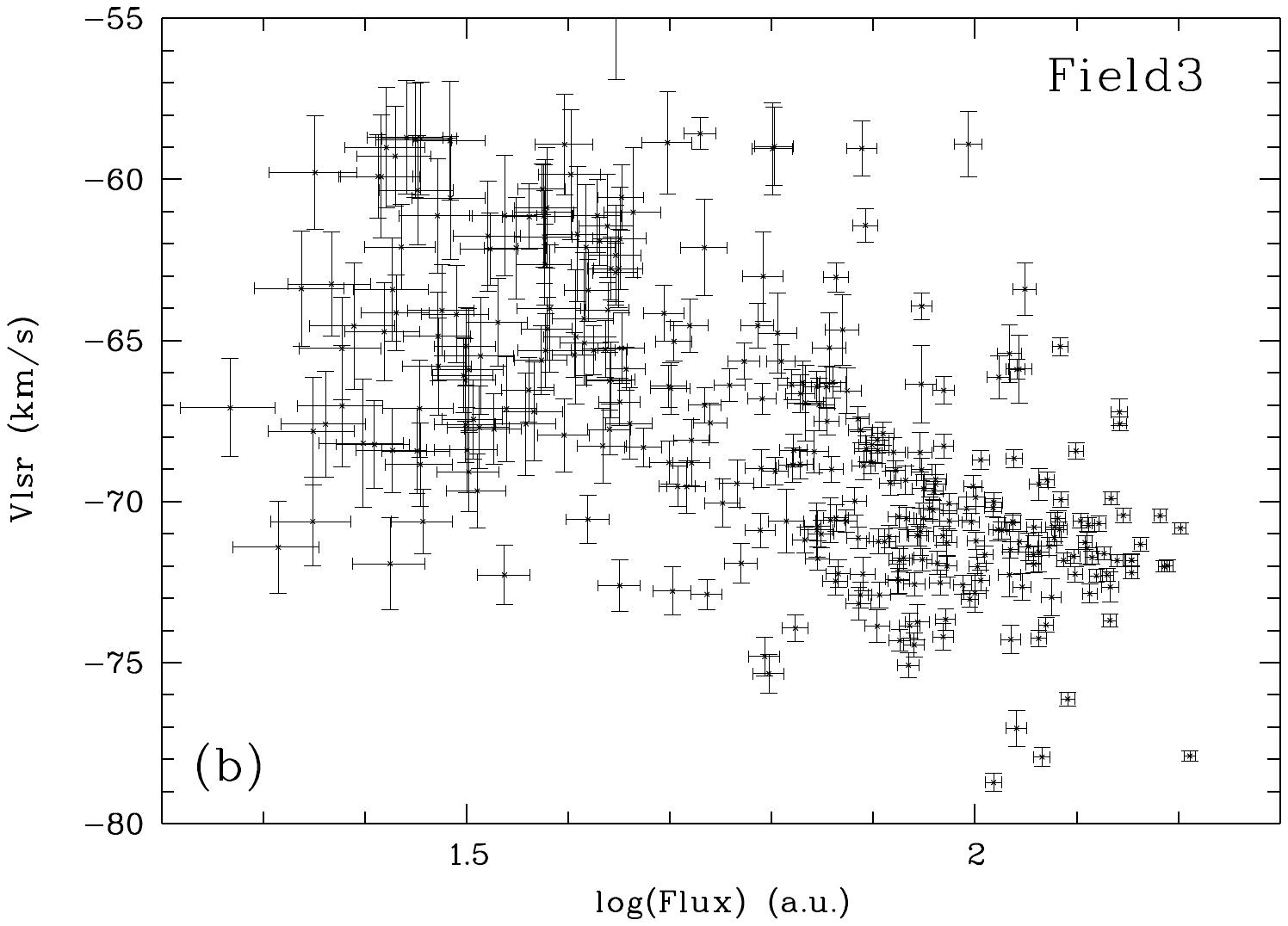} \includegraphics[scale=0.243,angle=-90,clip, viewport = 46 71 575 788]{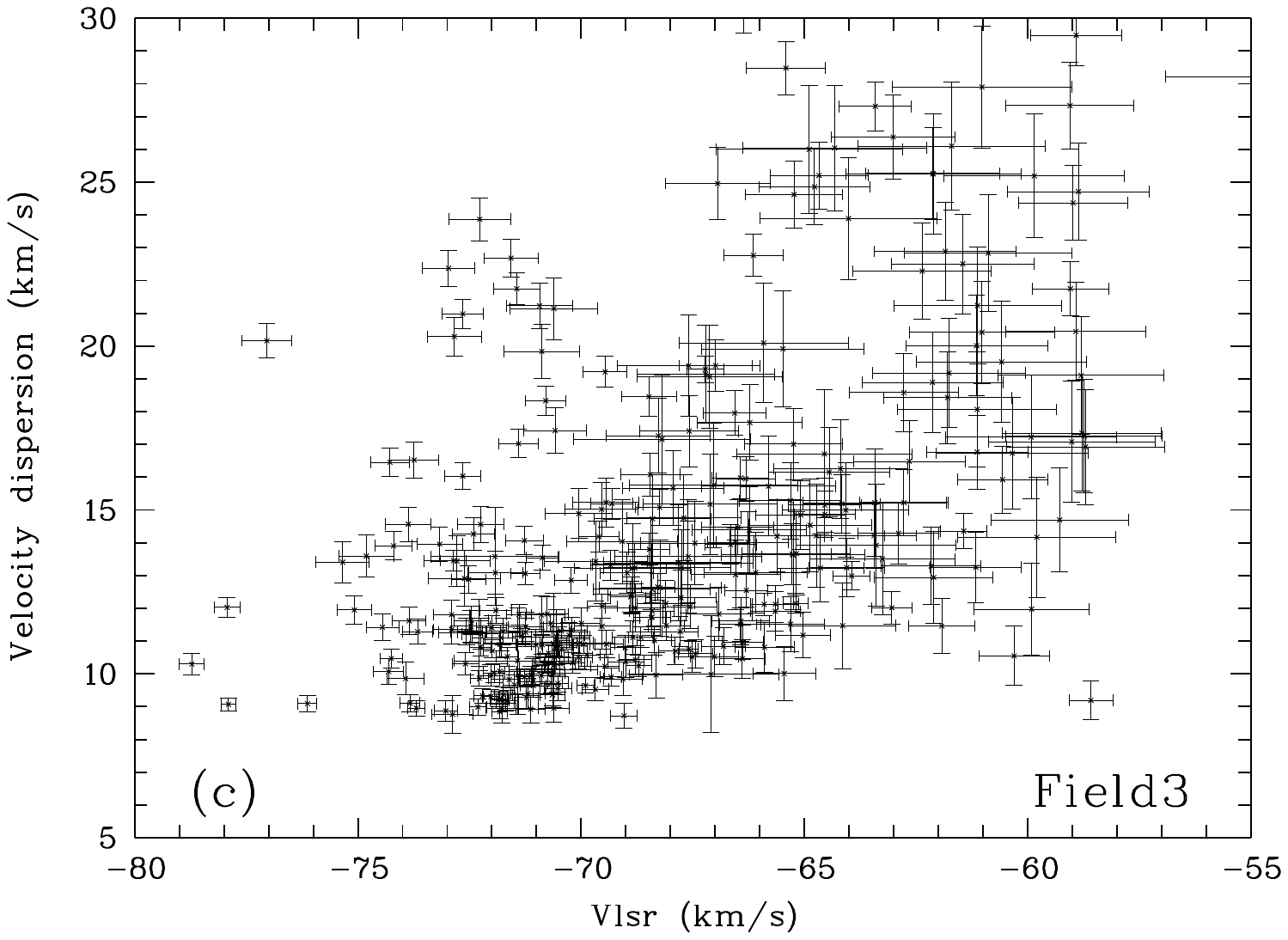}
\includegraphics[scale=0.243,angle=-90,clip, viewport = 46 71 575 788]{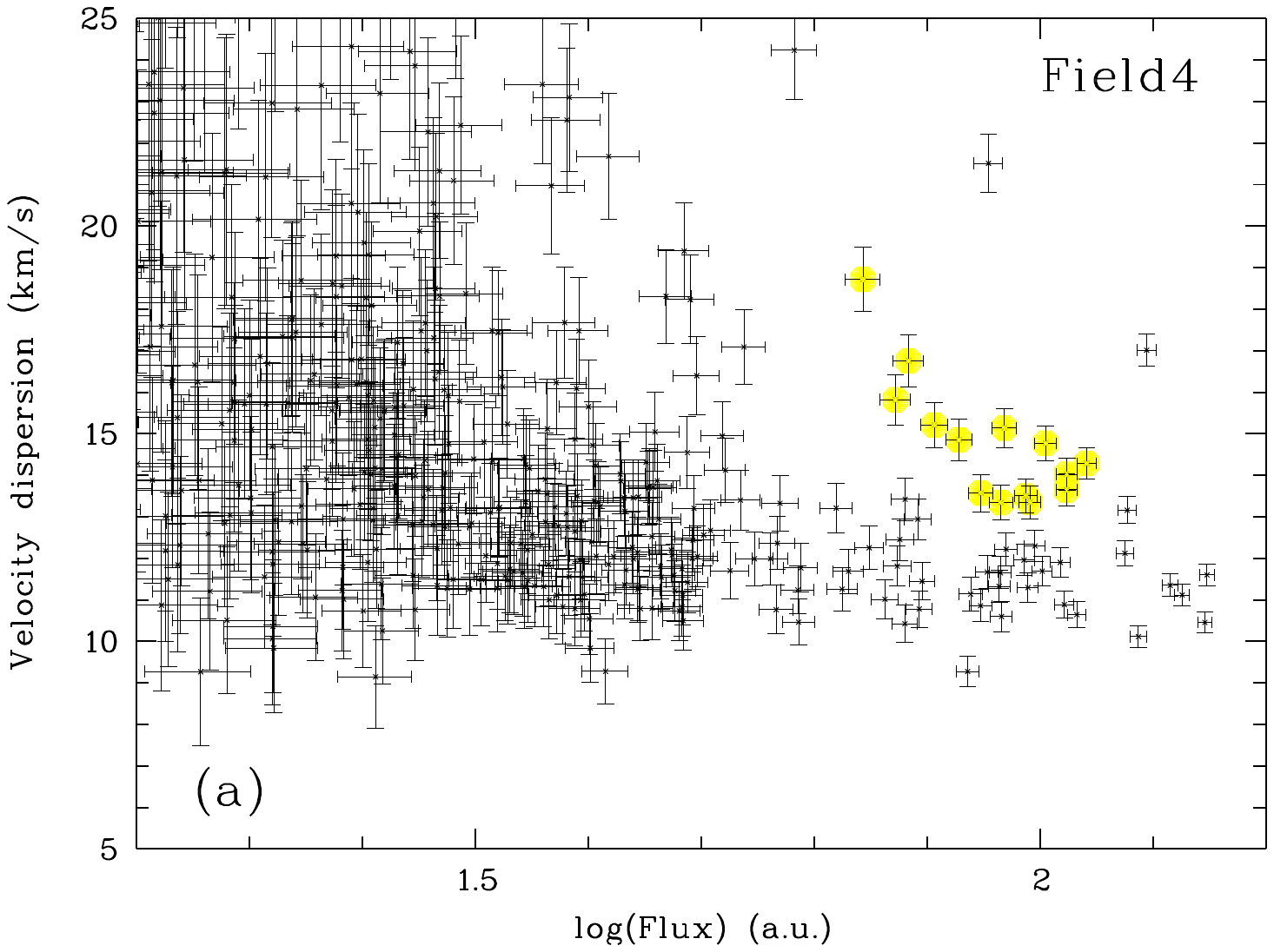} \includegraphics[scale=0.243,angle=-90,clip, viewport = 46 60 575 788]{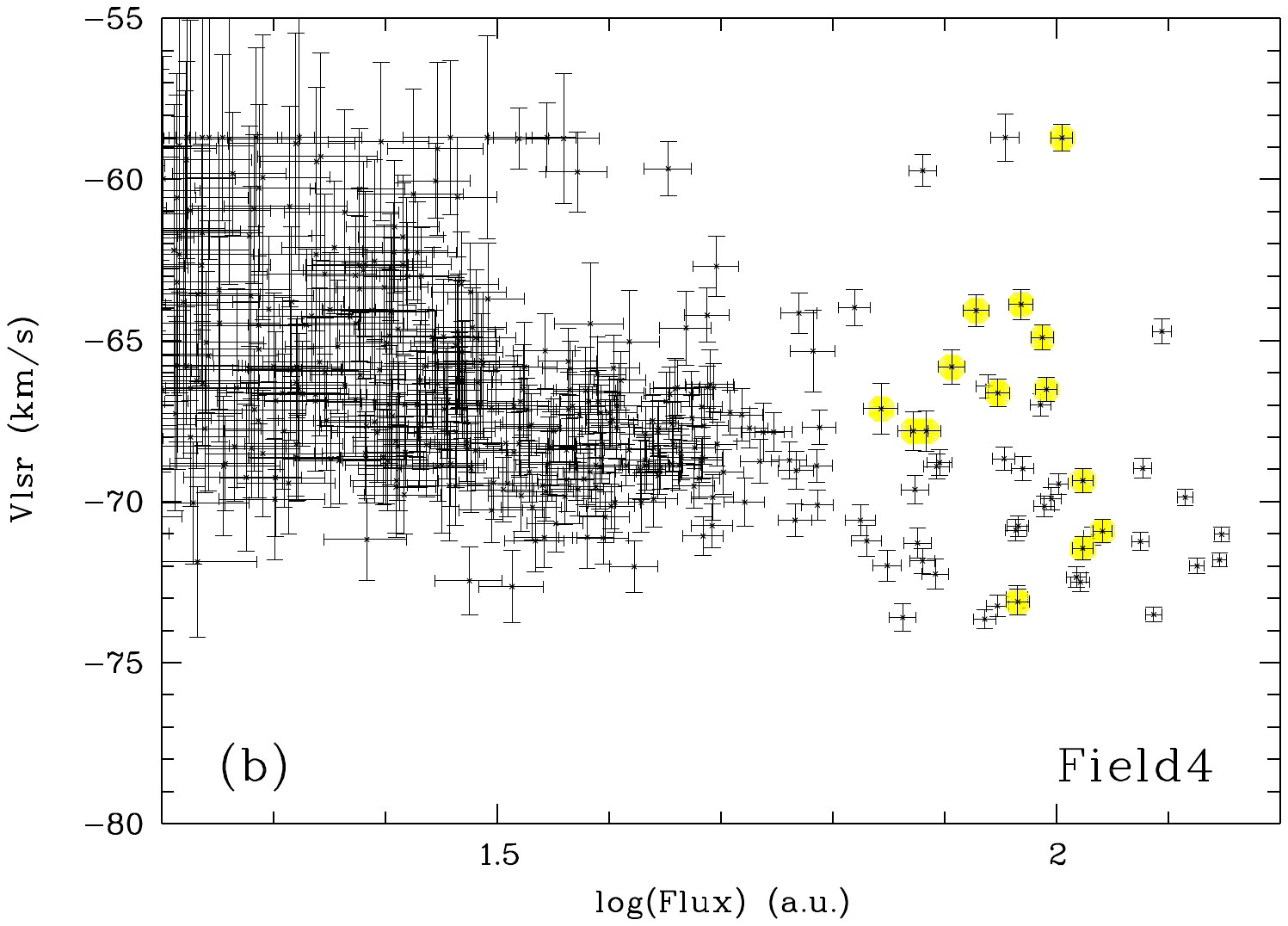} \includegraphics[scale=0.243,angle=-90,clip, viewport = 46 71 575 788]{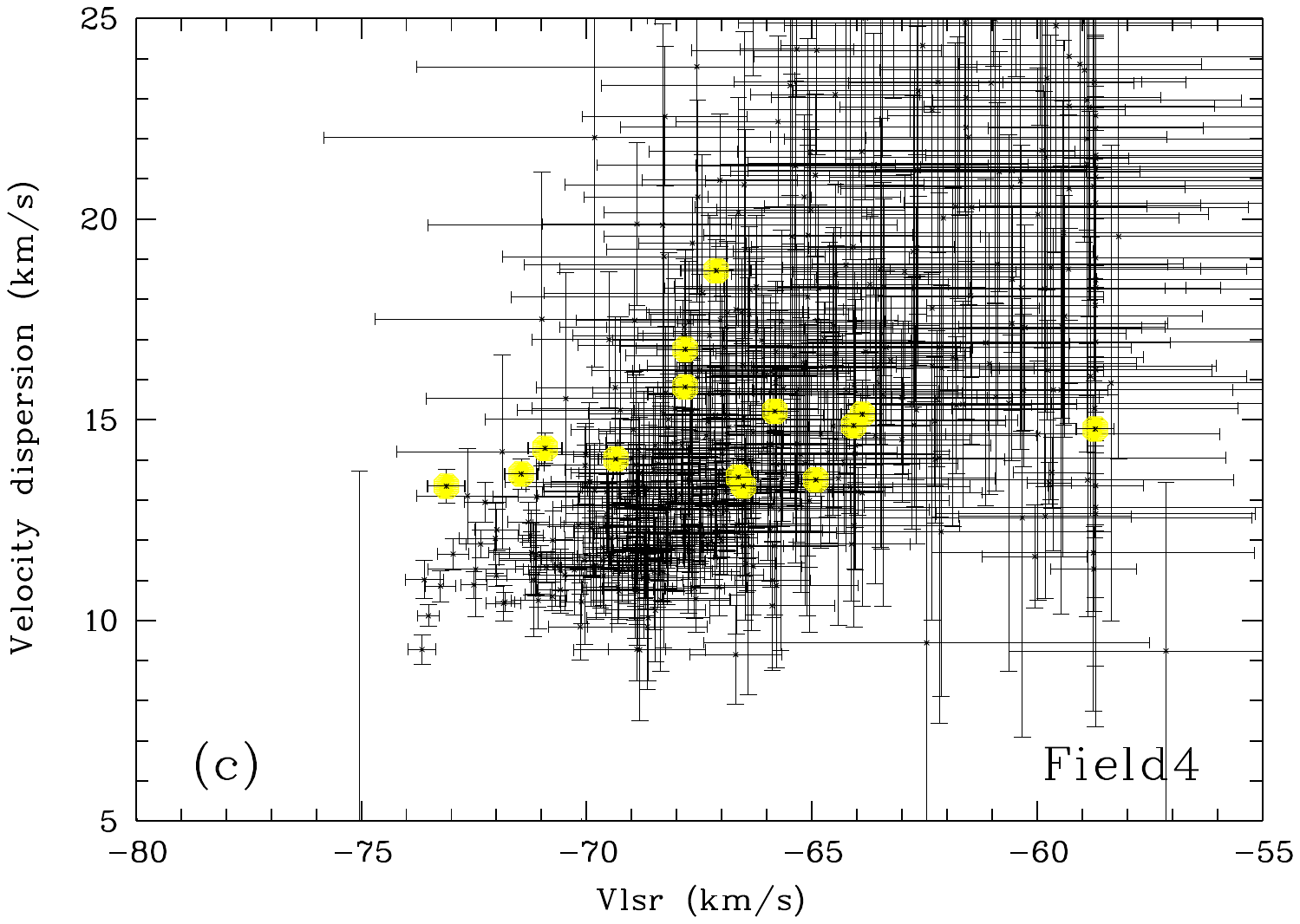}
\includegraphics[scale=0.243,angle=-90,clip, viewport = 46 71 575 788]{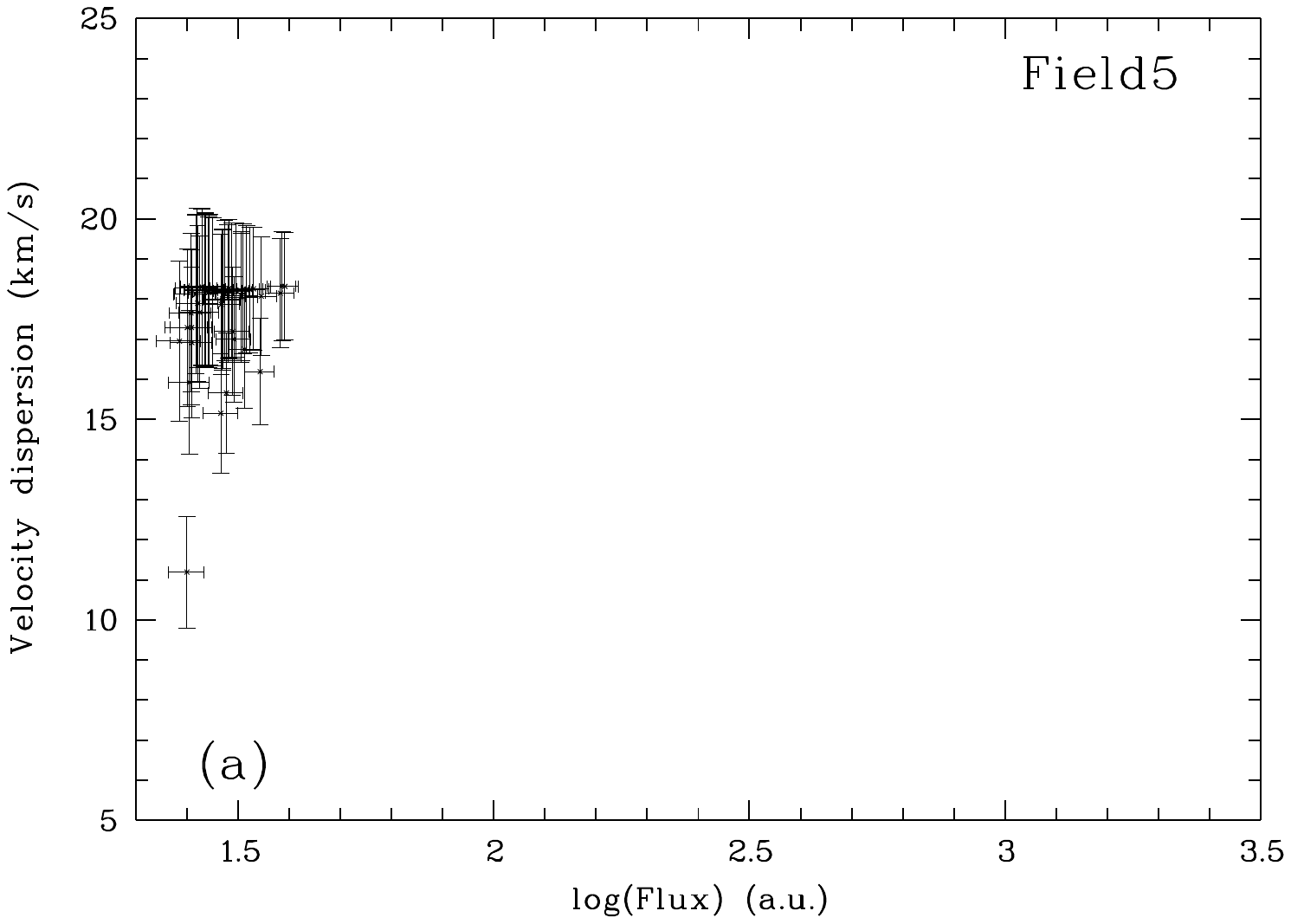} \includegraphics[scale=0.243,angle=-90,clip, viewport = 46 60 575 788]{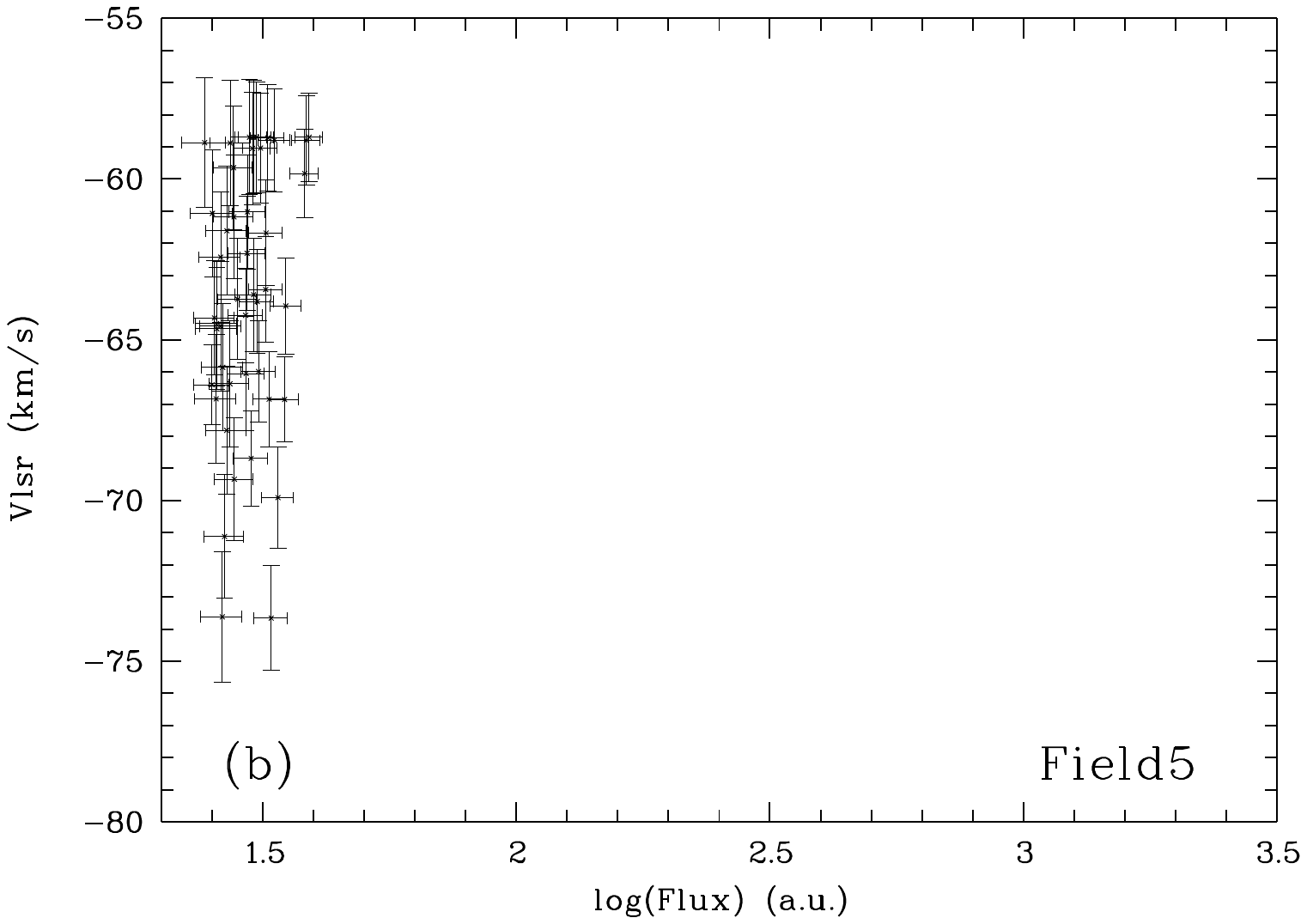} \includegraphics[scale=0.243,angle=-90,clip, viewport = 46 71 575 788]{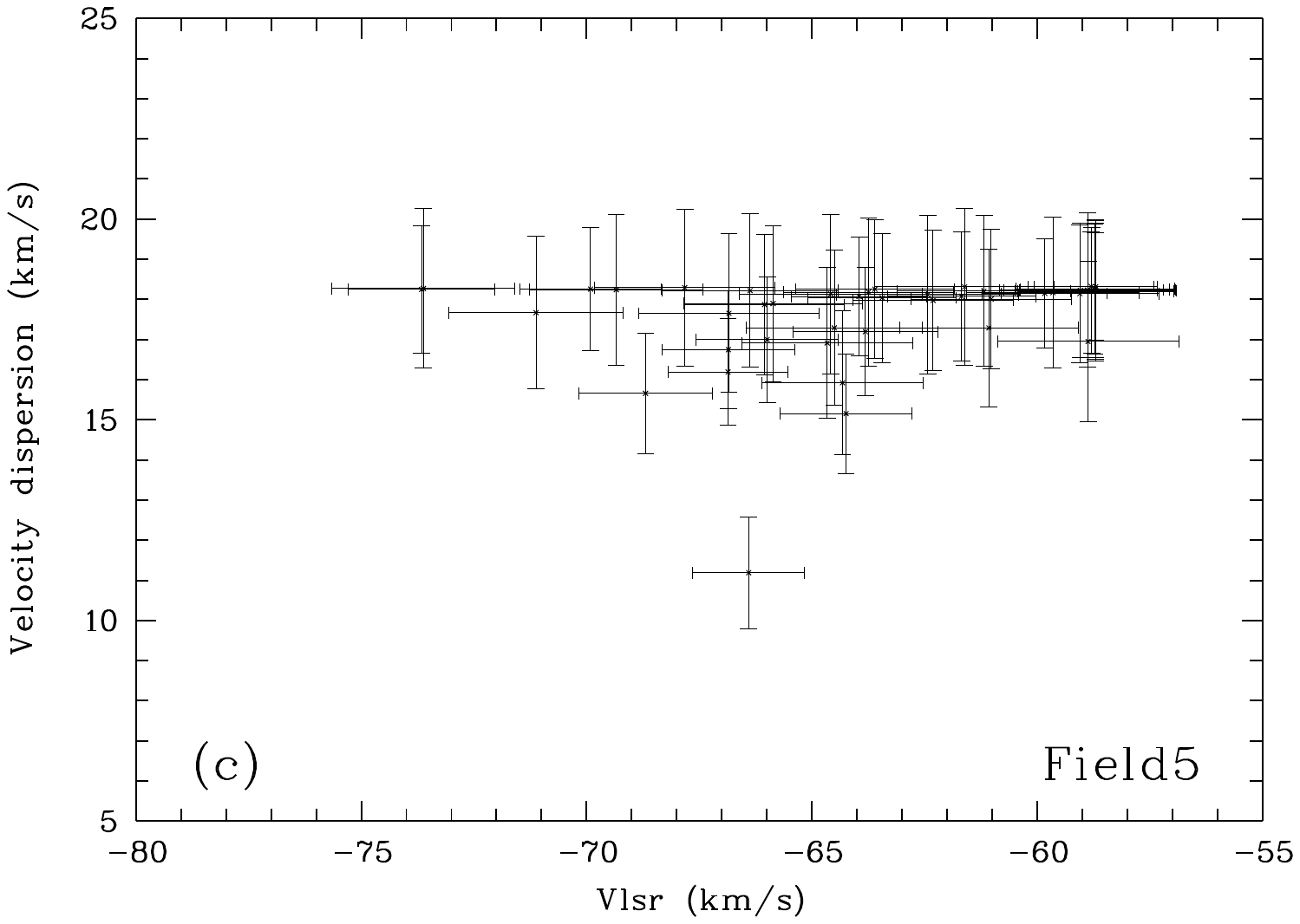}
  \caption{\label{kindiag} Kinematic diagrams for Fields 1 to 4: (a) $\sigma$ $-$ log(Flux)), (b) V$_{LSR}$ $-$ log(Flux) and (c) $\sigma$ $-$ V$_{LSR}$. The points highlighted in yellow, red, blue, and magenta identify groups discussed in the text.}
\end{figure*}

Non-thermal motions in \HII~regions are generally attributed to turbulence or large-scale velocity gradients. These velocity gradients are often associated with the ``champagne'' effect (\citealt{Franco90}) or shell expansion due to strong stellar winds. In extra-galactic giant \HII~regions, the study of the kinematic properties of the ionized gas is performed through the analysis of a set of kinematic diagrams (e.g., \citealt{Munoz-Tunon96}, \citealt{Bordalo09}, \citealt{Moiseev12}). These diagrams include the line intensity (logarithm of the line flux) versus velocity dispersion ($\sigma$ $-$ I), the radial velocity versus the intensity (V$_{LSR}$ $-$ I), and $\sigma$ $-$ V$_{LSR}$.

In these plots, $\sigma$ represents the non-thermal velocity dispersion. It is calculated from the line FWHM (previously corrected for the instrumental width) and corrected for the natural ($\sigma_{nat}$ = 3 km s$^{-1}$, \citealt{ODELL88}) and thermal (calculated here adopting a temperature for the region of 7890 K, \citealt{Luisi16}) widths. This temperature corresponds to a thermal sound speed of c$_{s}$ = 10.3 km s$^{-1}$ in the HII region. 

\cite{Munoz-Tunon96} indicate that large $\sigma$ values are expected at positions in the nebulae where loops, rings, shells, etc., are present. Specifically, shells (\citealt{Munoz-Tunon96}, \citealt{Moiseev12}) and outflows (\citealt{CruzGonzalez07}) are expected to draw inclined bands in the $\sigma$ $-$ I diagram, while \HII~regions should draw horizontal bands (\citealt{Munoz-Tunon96}, \citealt{Moiseev12}). Low-density turbulent interstellar medium should dominate at low intensity.

In the $\sigma$ $-$ V$_{LSR}$ diagram, a correlation between V$_{LSR}$ and $\sigma$ indicates any systemic motion with a significant component in the line of sight (such as champagne flows), while a random distribution may better represent isotropic expansion or turbulence (see \citealt{Bordalo09}). In the V$_{LSR}$ $-$ I diagram, a well-defined vertical band indicates local expansion of the gas and can be used to estimate the expansion velocity.

In this context, we constructed the kinematic diagrams (Figure \ref{kindiag}) for the five fields independently. We also retained the data with 6 km s$^{-1}$ $< \sigma <$ 30 km s$^{-1}$~and with peak intensity greater than 0.15 (arbitrary unit). Additionally, because observations are not flux calibrated and not corrected for the airmass, we do not plot the data for all the fields in the same diagram.

In Field 5, the number of points is low and due to its low emissions, this field is more affected by uncertainties. However, the data-point distributions are fairly consistent with those of a low-density warm interstellar medium.

In Fields 1 and 3, we clearly identify two groups of profiles. One group with $\sigma$ larger than 15 km s$^{-1}$ (supersonic) (Figures \ref{kindiag}-Field1-a and \ref{kindiag}-Field3-a) and with V$_{LSR}$ around $-$62 km s$^{-1}$ (Figures \ref{kindiag}-Field1-b and \ref{kindiag}-Field3-b), and the other following a horizontal band centered around $\sigma$ = 11 km s$^{-1}$, with a V$_{LSR}$ rather around $-$71 km s$^{-1}$. The first group originates mainly from the peripheral diffuse emission located to the east and north of Field 1 and to the north and on either side of the arch in Field 3. The second group clearly follows the horizontal band expected for \HII~regions.

Kinematic diagrams help us identify points that deviate from the main set. Nevertheless, uncertainties, especially at low signal-to-noise, can affect the inferred significance of these deviations and undermine the reliability of selecting points nearest the set.

\begin{figure}[]
\begin{center}
\includegraphics[scale=0.45, viewport = 35 176 536 585,clip]{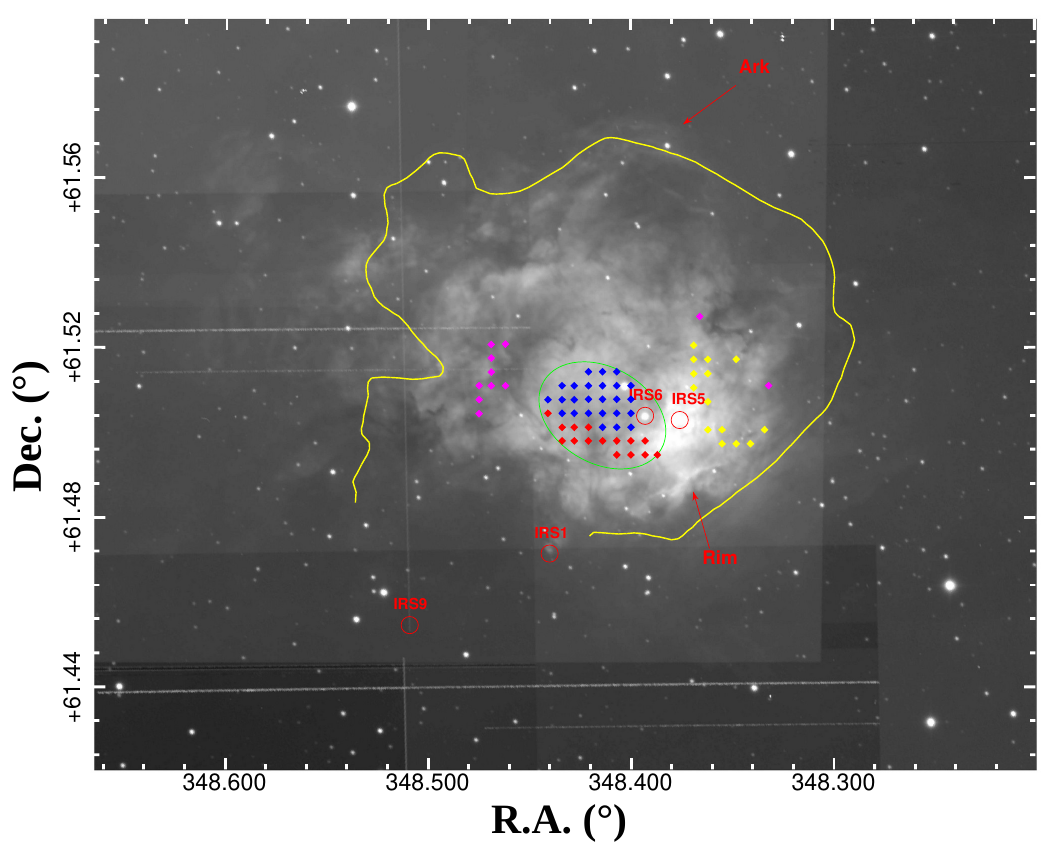}
  \caption{\label{kiniphas} IPHAS-H$\alpha$ image of NGC 7538. The coordinates are in degrees. This image allows us to better see the extinction features than in Figure \ref{ima1}.
   The yellow line and red names are similar as in Figure \ref{ima1}. The yellow, red, blue and magenta symbols are discussed in the text. The green ellipse (positioned and oriented by hand) models the opening of a conical outflow.}
\end{center}
\end{figure}

For Field 2, which covers the central part of NGC 7538, we can delineate, on Figure \ref{kindiag}-Field2(b), a group of points that stand out from the others (highlighted in blue) at V$_{LSR}$ 
around $-$64 km s$^{-1}$ but with a $\sigma \sim$ 10.5 km s$^{-1}$. They are located in the central part of the \HII~region (Figure \ref{kiniphas} $-$ blue symbols) and underline a local 
maximum of velocities (since they are in this area between $-$60 and $-$67 km s$^{-1}$). On Figure \ref{kindiag}-Field2(a), we can also delineate an inclined feature (traced by the red-
highlighted points) characteristic of a shell or an outflow. These points trace an elongated feature (Figure \ref{kiniphas} $-$ red symbols) of about 1.8 \arcmin (1 pc) located on the southern edge 
of the previous feature, but with a mean V$_{LSR}$ of $-$70 km s$^{-1}$ and a $\sigma \sim$ 16 km s$^{-1}$. At intermediate intensity, we note a vertical feature (traced by the magenta-
highlighted points) composed of points with a mean V$_{LSR}$ of $-$64 km s$^{-1}$ and located mainly in the eastern extinction features (Figure \ref{kiniphas} $-$ magenta symbols) between 
IRS6 and two cone-shaped rim-like structures (with stars at their tip) pointing towards IRS6 (\citealt{Sharma17}). This could then trace the presence of a quite active interaction between the 
\HII~region/IRS6 and the eastern PDR.

For Field 4, which covers mainly the diffuse emission to the west side of NGC 7538, most of the profiles show $\sigma$ larger than 10 km s$^{-1}$. The inclined feature (traced by the points highlighted in yellow, in Figure \ref{kindiag}-Field4-a) is drawn by profiles at the western edge of the \HII~region (Figure \ref{kiniphas} $-$ yellow symbols). This reflects a quite abrupt increase of $\sigma$ while the line of sight moves from the \HII~region to the diffuse emission, with a significant V$_{LSR}$ spread between $-$59 and $-$73 km s$^{-1}$.

Finally, the analysis of the $\sigma$ $-$ V$_{LSR}$ diagrams (Figure \ref{kindiag}-c) shows a clear trend only for Fields 3 and 4, underlining a gradient of about $-$15 km s$^{-1}$. For Fields 1 and 2, we note once more two groups of data-points: the first with $\sigma$ between 8 and 12 km s$^{-1}$and with V$_{LSR}$ between $-$62 km s$^{-1}$ and $-$75 km s$^{-1}$, and the other with larger $\sigma$ and a less negative velocity range (-58 km s$^{-1}$ and $-$64 km s$^{-1}$).

In summary, we see that the H$\alpha$ emission from beyond the region or from extinction dark features has a greater $\sigma$ and velocities between $-$58 and $-$64 km s$^{-1}$. The study also reveals a local maximum of velocities (mean velocity $-$63 km s$^{-1}$) at the center of the region, to the south of which an extended structure of 1.8\arcmin~is noticed with more negative velocities (-70 km s$^{-1}$), even though no specific structure is seen on the H$\alpha$ image. If we adopt a basic model where this would be the circular section of an inclined conical outflow, we trace an ellipse around these structures (green ellipse in Figure \ref{kiniphas}). We then can estimate a position angle of around 60\degree, which makes the cone's axis point back towards IRS1. Based on the ellipse's axes length, we estimate a diameter of 1.5~pc and an inclination of 45\degree~(respectively to the line of sight), resulting in an ejection velocity of around $-$89 km s$^{-1}$ (adopting V$_{LSR}$ = $-$63 km s$^{-1}$ as the central velocity) and an opening angle of 7\degree~(adopting V$_{LSR}$ = $-$70 km s$^{-1}$ as the edge velocity).

\subsection{Structure function analysis}
\label{sf}

The spectral information available through the region allows us to extract two important parameters: the mean width of the line profile, which gives the velocity dispersion of the emitting gas along the line of sight ($\sigma_{kin}$), and the mean standard deviation ($\sigma_{c}$) of the peak velocity over the region.

\begin{figure}[]
\includegraphics[scale=0.18,clip]{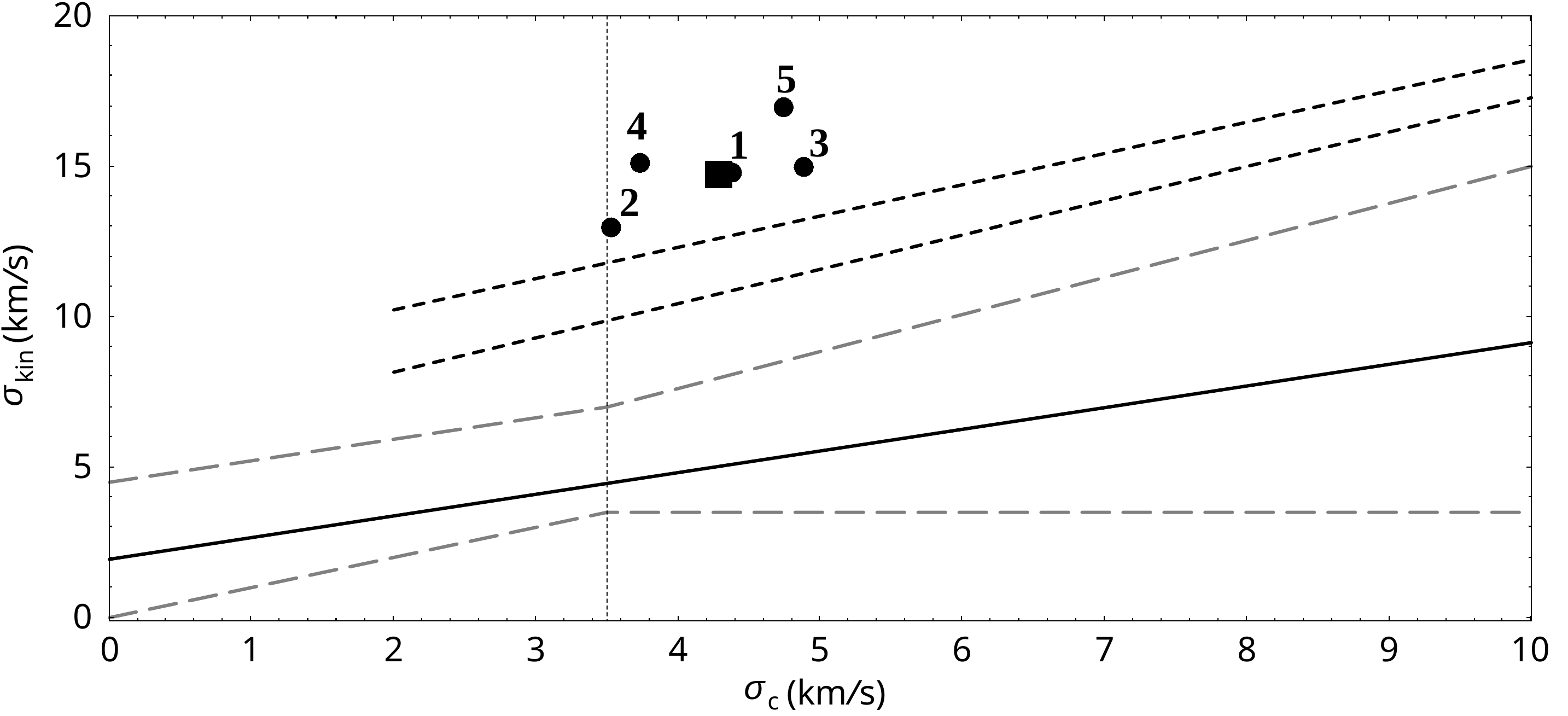}
\caption{\label{lagroisp} The line of sight dispersion $\sigma_{kin}$ versus the mean standard deviation $\sigma_{c}$ diagram. Fields 1 to 5 are plotted with filled circles and the filled square 
shows their average. The two dashed lines are from \cite{Lagrois11} and correspond to the fits applied to M33 \HII~region samples. The gray dashed lines delineate the spread of the models 
results around the solid line which indicates the linear regression applied to the whole sample models (with different angle of inclination). The vertical line indicates the rough separation 
between the wind-less Champagne flow models (on the left) and the Champagne flow perturbed by the expanding wind-blown bubble models (on the right).}
\end{figure}

$\sigma_{c}$ and $\sigma_{kin}$ are expected to be correlated with large-scale champagne flows and velocity gradients, photo-ablation effects, colliding flows, or turbulent motions, resulting in large non-thermal widths (\citealt{Lagrois11}). Based on hydrodynamical, non-turbulent models of expanding \HII~regions from \cite{Arthur06}, \cite{Lagrois11} show the expected location in the $\sigma_{kin}$ $-$ $\sigma_{c}$ plot (see their figure 6) for Champagne models perturbed or not by a stellar wind-blown bubble and for backward and forward flows (with tilt angles between 95\degr~and 175\degr). As observed by \cite{Lagrois11} for \HII~regions in M33, we find here that the different fields of NGC~7538 (Figure \ref{lagroisp}) also fall well above the simulated values, suggesting that there is a large part of turbulence or additional large-scale ordered motions.

\begin{figure}[h]
  \includegraphics[scale=0.16,clip]{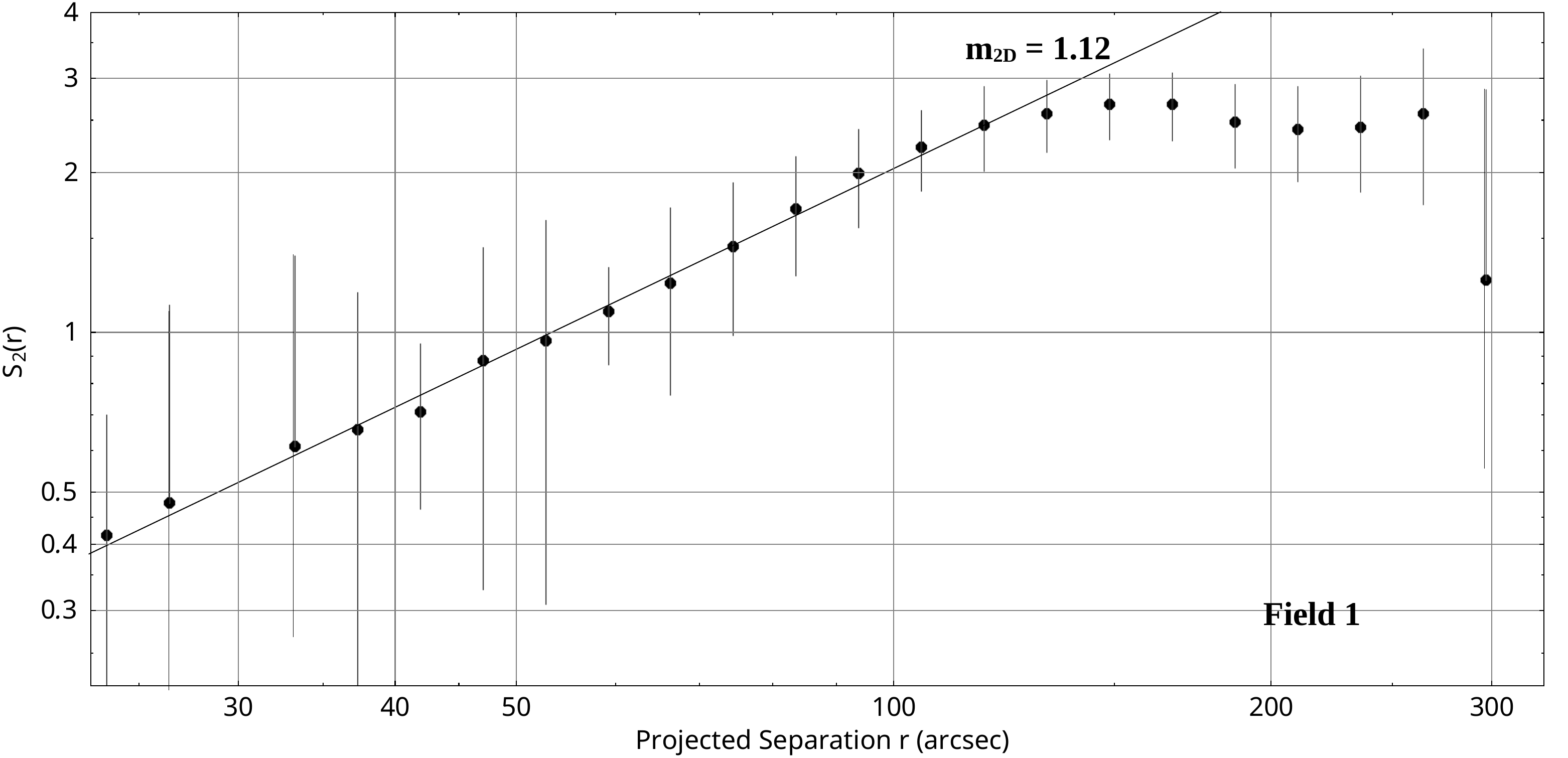}
    \includegraphics[scale=0.16,clip]{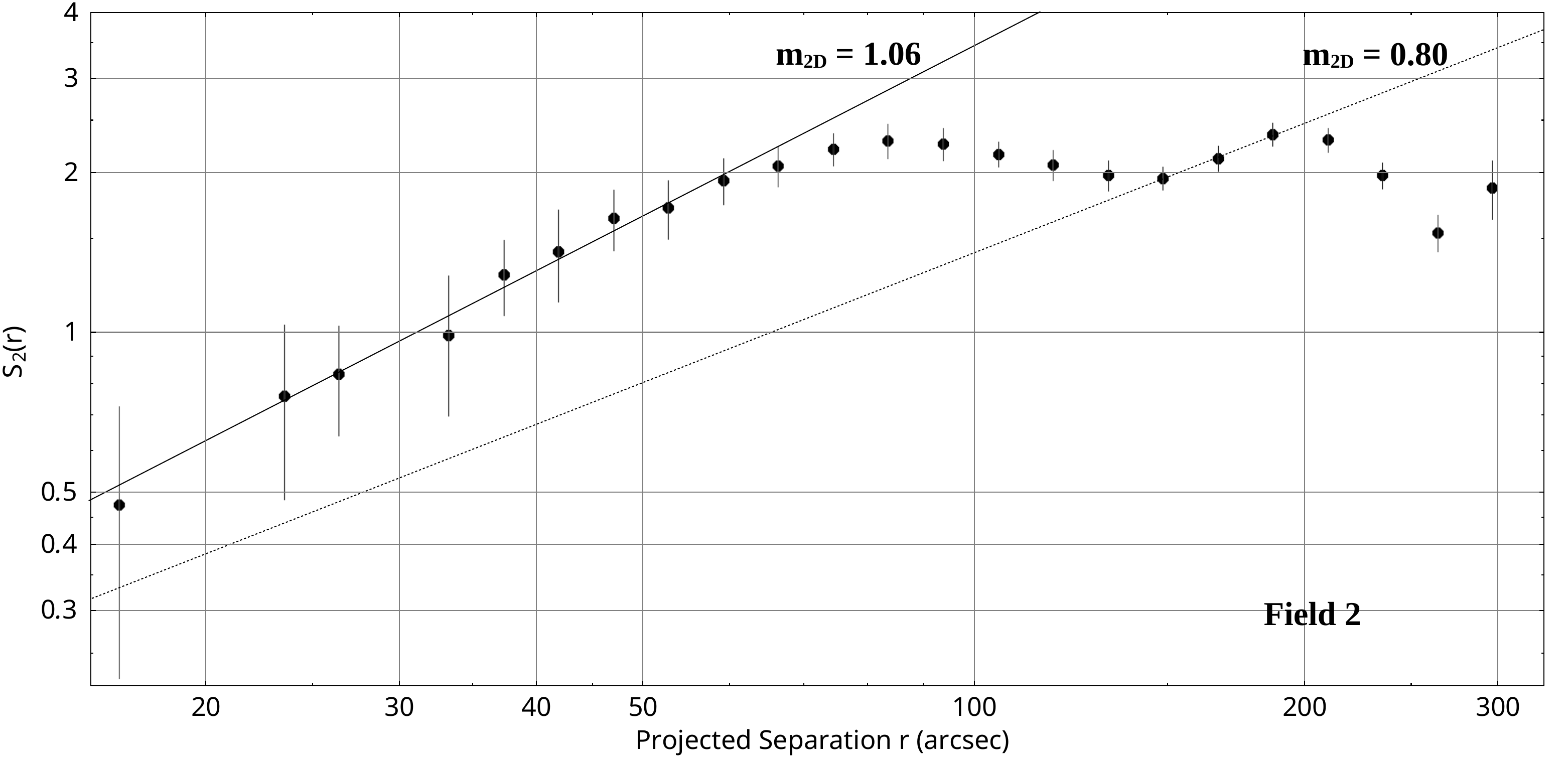}
     \includegraphics[scale=0.16,clip]{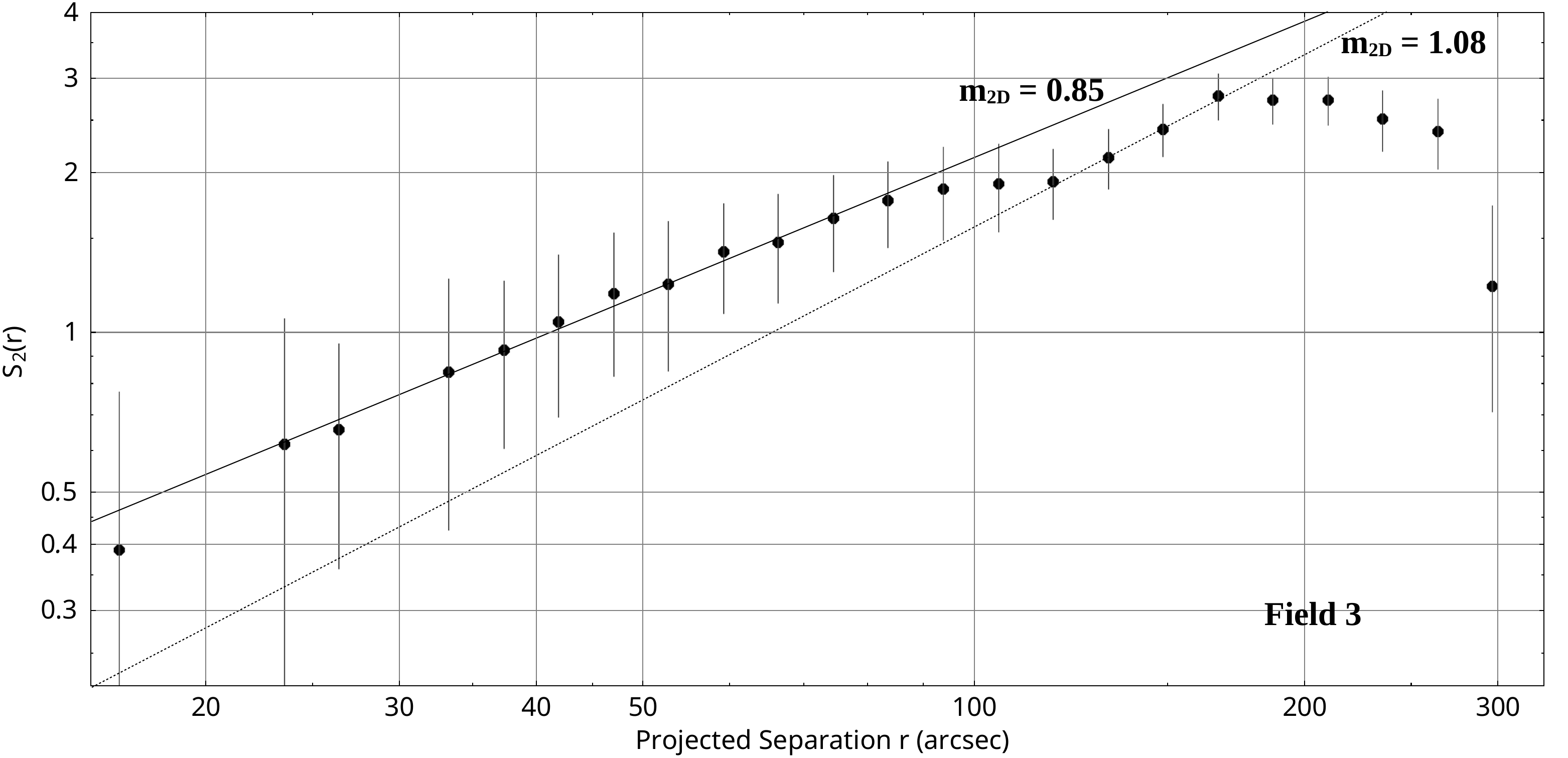}
      \includegraphics[scale=0.16,clip]{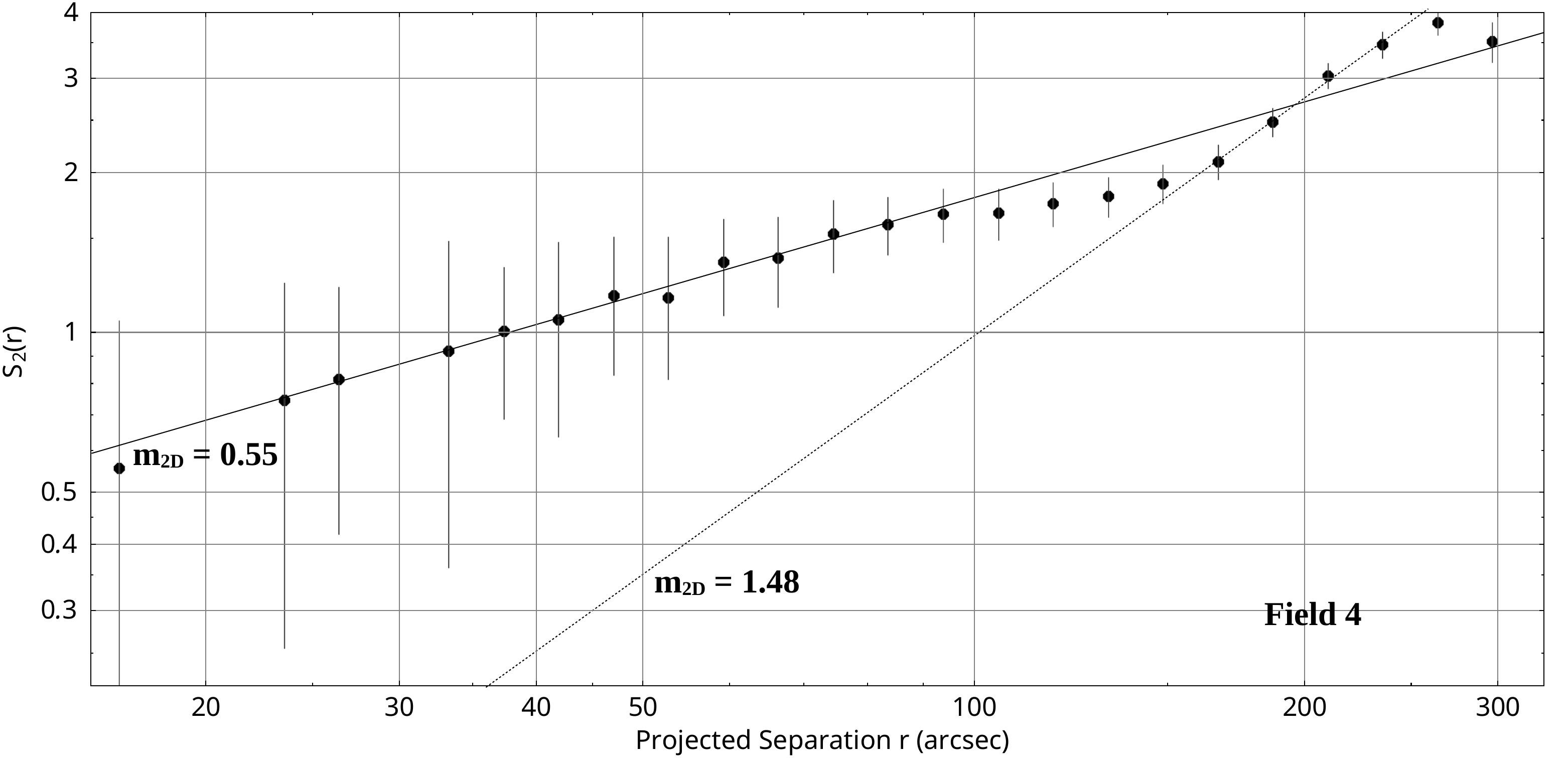}
     \includegraphics[scale=0.16,clip]{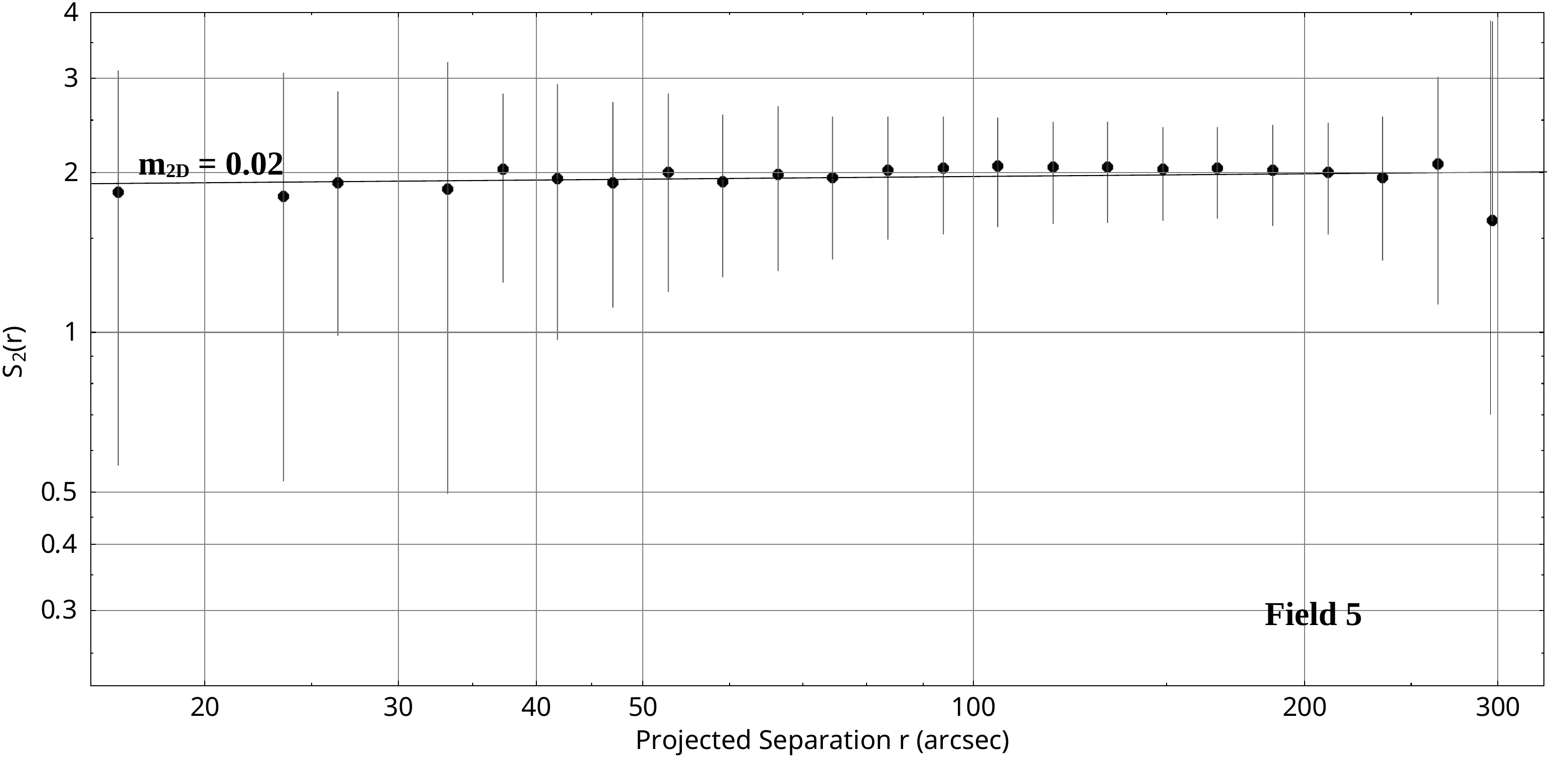}   
  \caption{\label{strucf2} Log-log plot of the second order structure function S$_{2}$(r) for the five fields. The fitted power-law are shown and their indices are indicated in each panel.}
\end{figure}

Following \cite{Arthur16}, \cite{Miville95} and \cite{GarciaVasquez23}, to characterize the spatial scale of this turbulence, we build the second-order structure function, S$_{2}$ (Figure \ref{strucf2}), of the velocity centroid (using equations 1 to 4 in \citealt{Arthur16}), which measures the variation of the velocity as a function of the plane-of-sky separation (r).

From molecular cloud simulations, \cite{Chira19} show that the velocity structure function (VSF) cannot be described by a single power-law relation over the entire range of r and that only the small and intermediate ranges may be represented by a common power-law. For \HII~regions, \cite{Medina14} and \cite{GarciaVasquez23} observe that they exhibit a rising VSF with a power-law index, m$_{2D}$, between 0.5 and 1 at the smallest scales, which transitions to a flat structure function with m$_{2D} \sim 0$ at larger scales.

The relation between the three-dimensional power-law index, m$_{3D}$, and the two-dimensional one, m$_{2D}$, depends on the projection smoothing effect. If the emission is sheet-like, hence 
for projected separations larger than the characteristic line-of-sight depth of the emitting gas, m$_{2D} \approx$ m$_{3D}$, while at smaller separations than this, m$_{2D} =$ m$_{3D} + 1$ 
(e.g., \citealt{Arthur16}, \citealt{Medina14}). Then the slope will be affected by geometry and systematic motions (\citealt{Medina14}). For example, a more uniform density distribution will result in 
a steeper slope. Conversely, if the emission originates from a layer close to an ionisation front, or if there is significant density variation, the slope becomes smaller. Similarly, when the viewing 
angle is aligned with the flow direction, systematic anisotropic flows (e.g. champagne flows) can affect the slope, making it steeper at large scales. 

Because the H$\alpha$ emission is expected to occupy the entire volume of the \HII~region and the optical depth varies from point to point over the face of the region, it is difficult to determine if 
the NGC 7538 region is sheet-like or not. Thus, a theoretical m$_{3D}$ provides us with a range for m$_{2D}$. Indeed, from a theoretical point of view, \cite{Medina14} recall that m$_{3D}$ = 2/3 
(0.66 $<$ m$_{2D}$ $<$ 1.66) in the case of incompressible, homogeneous (Kolmogorov-like) turbulence, but intermittency and compressibility can increase this to m$_{3D}$ $\sim$ 0.8 (0.8 
$<$ m$_{2D}$ $<$ 1.8) in subsonic regime and up to m$_{3D}$ = 1 (1 $<$ m$_{2D}$ $<$ 2) for compressible, shock-dominated, and supersonic turbulence (\citealt{Schmidt08}). However, in 
the case of an \HII~region, the medium is mildly supersonic and should show an intermediate configuration. In this context, \cite{Medina14} find, from \HII~region simulations, that, due to the 
complex  geometry and varying line-of-sight depths in real \HII~regions, at late times, m$_{2D}$ should be between 0.2 (in the large-scale limit) and 1.2 (in the small-scale limit) while 
\cite{Arthur16} and \cite{GarciaVasquez23} find, from H$\alpha$ observations, m$_{2D}$ = 1.17 and  m$_{2D}$ = 1.03, on average, for the Orion nebula and from a sample of ten \HII~regions, 
respectively.

The analysis of S$_{2}$ (Figure \ref{strucf2}) for the different fields shows, with the exception of Field 5, a power-law increasing curve followed by a more or less pronounced plateau, underlining decorrelation. The S$_{2}$ break indicates either the size of the largest turbulent eddies from which the cascade of turbulent kinetic energy could take place (e.g., \citealt{Castaneda88}, \citealt{Lagrois11b}) or the value at which the depth of the line of sight equals this scale (\citealt{Medina14}). In this context, we note that m$_{2D}$ is around 1 for Fields 1 to 3, in accordance with what has been found for other \HII~regions by \cite{GarciaVasquez23}. It is also more consistent with some compressible and shock-dominated regime. In addition the shallower slope in Field 3 suggests that it is more strongly dominated by shocks than Fields 1 and 2. For Field 4, which mainly probes the ionized gas in the direction of the molecular cloud, m$_{2D}$ is 0.55, more typical of Kolmogorov-type turbulence. The correlation length is between 1.02 and 1.46 pc (78\arcsec~- 112\arcsec) for Fields 1, 3, and 4, while it is $\sim$0.72 pc (55\arcsec) for Field 2.

Field 5 shows a flat S$_{2}$, underlining homogeneous and random fluctuations typical of thermal instability (\citealt{Mohapatra22}). This field probes the warm interstellar medium surrounding the \HII~region.

At larger scales, Fields 2, 3, and 4 exhibit a second power-law, which is usually attributed either to periodicity or large-scale gradients in the velocity field (\citealt{Lagrois11b}, \citealt{GarciaVasquez23}). In this context, for Field 2, the wavy aspect of the curve after 70\arcsec~would better suggest a periodic oscillation (with a wavelength of $\sim$1.9 pc). For Fields 3 and 4, as these fields straddle the edges of the \HII~region, the second power law could better be interpreted as a large-scale gradient of the velocity field crossing the PDR and characterizing the velocity contrast between the interior of the \HII~region and the outer, more diffuse emission.

From the extreme points which define the second power, we can estimate for each of these two fields the characteristic scale and velocity gradient ranges. These are between 1.9 $-$ 2.2 pc and 6.5 $-$ 7.8 km s$^{-1}$for Field 3 and 2.2 $-$ 3.1 pc and 5.3 $-$ 6.5 km s$^{-1}$for Field 4, respectively.

\begin{figure}[h]
  \includegraphics[scale=0.165,clip]{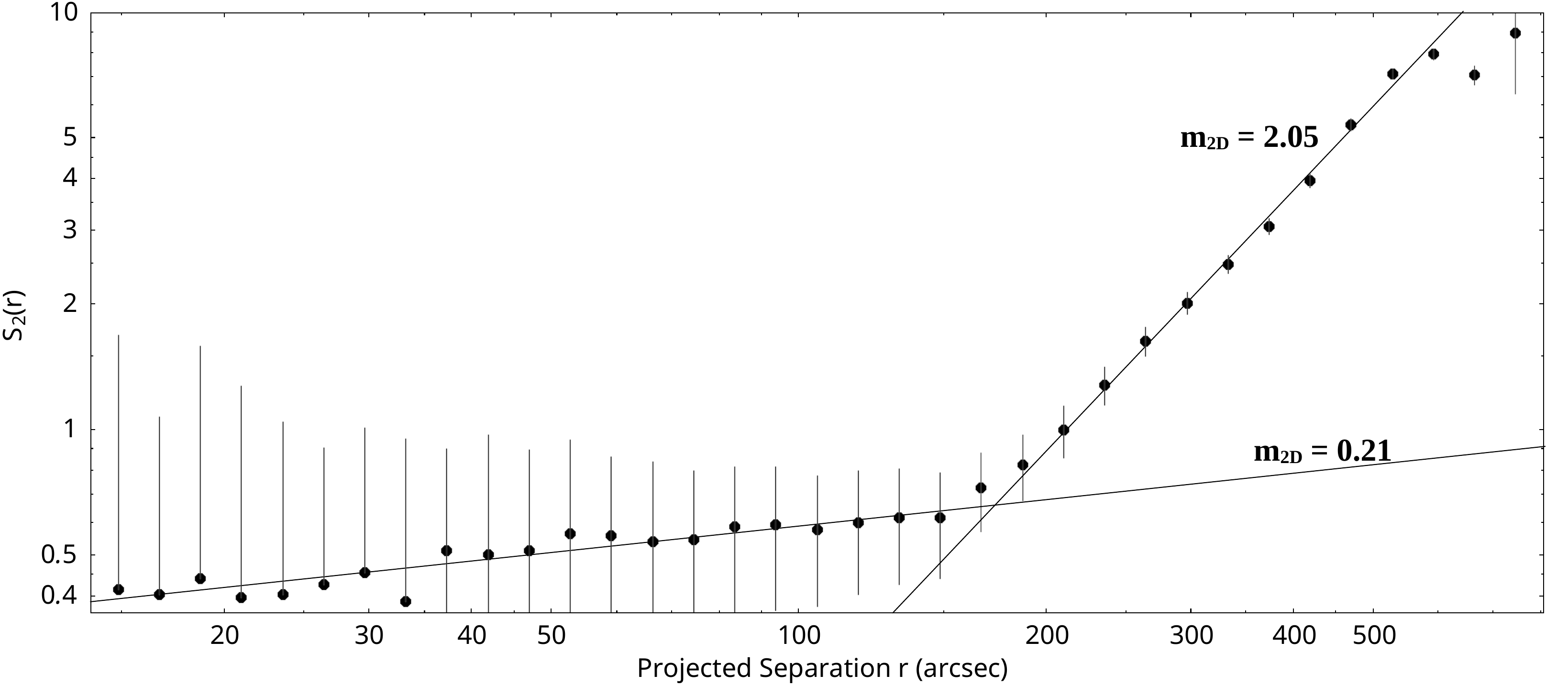}
  \caption{\label{strucf3} Log-log plot of the second order structure function S$_{2}$(r) for the full area.  The fitted power-law are shown and their indices are indicated in the panel.}
\end{figure}

To probe larger scales, we constructed the structure function for all fields combined (Figure \ref{strucf3}) from profiles with a higher peak intensity (larger than 0.5 in arbitrary unit) to ensure that the profiles used to build the structure function have a good signal-to-noise ratio. Since the slope of the structure function depends on the dominant source of turbulence, we expect the S$_{2}$ constructed using all fields to be dominated by large-scale, anisotropic motions (e.g. Champagne flow). This will make the slope steeper at large scales and shallower at small scales, as can be seen in Figure \ref{strucf3} where the first and second power-laws are shallower (m$_{2D}$ = 0.21) and steeper (m$_{2D}$ = 2.05), respectively, than for individual fields. The shallow slope of the first power law is consistent with \cite{Royer25} (who report a slope of 0.251 in their Figure 22 but for the [N\,{\sc{ii}}] line). The second power-law starts around 2.4 pc (186\arcsec), a similar value as for Fields 3 and 4. The large-scale motions are estimated to occur between $\sim$2.4 pc (with a velocity gradient of $\sim$7.9 km s$^{-1}$) and $\sim$6.9 pc (with a velocity gradient of $\sim$23 km~s$^{-1}$).

To summarize this section, we find a correlation length between 0.72 and 1.46~pc. In addition, in Fields 3 and 4, a typical velocity gradient of about 6.5 km s$^{-1}$ for scales around 2.2 pc is put in evidence, while in Field 1, this is not observed. These typical sizes and velocities correspond to the typical characteristics of bubbles observed by \cite{Beuther22} in NGC 7835. Field 2 is particular as a periodic oscillation with a wavelength of $\sim$1.9~pc is underlined. Large-scale motions are then estimated to occur between $\sim$2.4 pc (with a velocity of $\sim$7.9 km s$^{-1}$) and $\sim$6.9 pc (with a velocity of $\sim$23 km s$^{-1}$). Shock-dominated turbulence is suggested for Fields 1 to 3, while for Field 4, turbulence seems more Kolmogorov-like.




\section{Discussion}
\label{discsec}

Previous optical observations of NGC 7538 reported H$\alpha$ velocities ranging from $-$44 km s$^{-1}$to $-$71 km s$^{-1}$, with a mean velocity of $-$58.5 km s$^{-1}$ (\citealt{Deharveng79}). All ionic radial velocities are blue-shifted with respect to the parental molecular cloud (\citealt{Barriault07}). More specifically, \cite{Barriault07} showed that the ionized gas flows outward toward the observer, with several arguments supporting a Champagne model. More recently, \cite{Beuther22} identified multiple bubble-like structures in [CII] 158 $\mu$m Stratospheric Observatory for Infrared Astronomy (SOFIA) observations, with expansion velocities around 10 km s$^{-1}$and extreme blue-shifted velocities observed at approximately $-$70 km s$^{-1}$. \cite{Luisi16} measured the hydrogen radio recombination line at 8.7 GHz across the region, finding a mean velocity of $-$58.5 km s$^{-1}$, which closely matches the molecular cloud velocity (V$_{CO}$ = $-$56.1 km s$^{-1}$ from \citealt{Blitz82}). Recently, \cite{Royer25} conducted a study on the turbulence in NGC 7538, focusing on the [NII]6584\AA~line profiles. Their findings indicate that the turbulence in the velocity field is supersonic, ranging from 1.3 to 1.7 times the sound speed, and exhibits spatial intermittency, except at the rim location.
Additionally, they examined the VSF of the [N II] line. Given that the [N II] emission originates from a relatively thin layer, projection smearing is expected to be minimal. Consequently, their measured power-law index is directly representative of the three-dimensional index, m$_{3D}$. Their results show a power-law index decreasing from 0.57 in the diffuse part of the region to 0.4 in the rim (referred to as the filament region in their study). Their area of investigation corresponds mainly to our Field 2 for which we find a larger power-law index consistent with the findings of \cite{Arthur16}, who demonstrated that in M42 the [N II] VSF is shallower than the H$\alpha$ VSF.

The main results from Sections \ref{diag} and \ref{sf} suggest that the kinematics in the center of the region may be influenced by an outflow originating from IRS1. Additionally, we observe a velocity gradient of approximately 23 km s$^{-1}$ on the scale of the region and a gradient of around 6.5 km s$^{-1}$ on the scale of 2.2 pc. In Fields 1 to 3, the turbulence appears to be shock-dominated, while in Field 4, it exhibits a more Kolmogorov-like behaviour. However, the "$-$36 km s$^{-1}$" and "$-$92~km s$^{-1}$" components were not considered in these sections.

On Figure \ref{maps}, as noted by \citealt{Deharveng79} and \cite{Barriault07}, the H$\alpha$ emission is systematically blue-shifted relative to the velocity of its parental molecular cloud 
(V$_{\rm{CO}}$ = -56.1 km s$^{-1}$ from \citealt{Blitz82}), with no preferential direction, which is consistent with the expected behavior in a Champagne flow. Depending on whether we adopt 
the mean velocity (-67 km s$^{-1}$) or the extreme velocity (-77 km s$^{-1}$) of the main component (observed in Field 2), we estimate a flow velocity between 11 and 21 km s$^{-1}$. This is in 
agreement with the large-scale motion described in Section \ref{sf}.

As already noted in Section \ref{diag}, we observe that profiles originating from extinction-dark features or from outside the region (Field 5, north of Field 3, and most of Field 4) show velocities between $-$64 and $-$58 km s$^{-1}$, with $\sigma$ values larger than 15 km s$^{-1}$ (larger than the typical velocity dispersion observed in the Warm Interstellar Medium, \citealt{Haffner98}, \citealt{Madsen06}). In this context, the large amount of foreground dust features, seen as dark regions at the front of the region (e.g., at $\alpha$, $\delta$ = 348.457\degree, +61.516\degree~and 348.418\degree, +61.495\degree~in Figure \ref{kiniphas}), complicate the interpretation of the velocity and velocity dispersion maps. Additionally, no splitting of the H$\alpha$ line is observed, which excludes, at first glance, any wind-bubble structure. However, the intrinsic large width of the H$\alpha$ line makes it difficult to detect bubbles with expansion velocities between 6 and 14 km s$^{-1}$, as observed by \cite{Beuther22} from [CII] 158 $\mu$m SOFIA data.

We next examine specific structures visible in the H$\alpha$ image (Figure \ref{kiniphas}) and identified on the maps (Figure \ref{maps}). At the rim, no significant increase in $\sigma$ is observed, while the V$_{LSR}$ shifts from $-$72 to $-$66 km s$^{-1}$ from the inner to the outer regions along a cut perpendicular to the rim. The arc shows a possible velocity gradient, with V$_{LSR}$ between $-$77 and $-$66 km s$^{-1}$and $\sigma$ values between 13 and 17 km s$^{-1}$ from east to west. The brightest H$\alpha$ emission surrounding IRS5 has a mean velocity of $-$72 km s$^{-1}$ ($\sigma$ = 13.6 km s$^{-1}$) and is part of a larger semi-circular emission at V$_{LSR}$ = $-$71 km s$^{-1}$ ($\sigma$ = 14.5 km s$^{-1}$), with its main feature located between IRS1 and IRS6. This feature is consistent with the outflow model suggested in Section \ref{diag} and appear to be linked to a feature characterized by a high [OIII] (4959\AA + 5007 \AA) / H$\beta$ ratio, indicative of a high ionization parameter, as identified by \cite{Royer25} (see their figure 5-a). A similar kinematic pattern, already illustrated in Figure 3 of \cite{Dickel81}, was also observed by \cite{Barriault07} in the inner region and was attributed to the collision of two gas flows: one northwesterly flow directed at least partially toward the observer, and one southwesterly flow more inclined away from the observer.

In addition to the main velocity component, we observe two additional faint velocity components on either side of IRS6 (about 1.2') and along the semi-circular feature (Figure \ref{maps} - middle panels): a redshifted component around $-$35~km s$^{-1}$ and a blueshifted component around $-$92 km s$^{-1}$ (Figure \ref{maps} $-$ lower panels). On average, relative to the main component, the $-$35 km s$^{-1}$ component has an intensity of 10\%, a mean of $\sigma$ of 13 km s$^{-1}$, and a mean velocity difference of 37 km s$^{-1}$, while the $-$92 km s$^{-1}$ component has an intensity of 20\%, a mean $\sigma$ of 8 km s$^{-1}$, and a mean velocity difference of $-$22 km s$^{-1}$. These additional components make the profiles asymmetric and are typically attributed to photoevaporation, dust scattering, shell expansion, or outflows.

\cite{Henney05} showed that recombination lines formed in photoevaporation flows, such as H$\alpha$, are shifted relative to the neutral gas emission by about 10 km s$^{-1}$.
In this context, both the $-$35 km s$^{-1}$ and $-$92 km s$^{-1}$ components exhibit a significant velocity difference compared to the molecular cloud, making it unlikely they are due to photoevaporation.

\cite{Relano05}, \cite{Rozas06a}, and \cite{Rozas06b} observed low-intensity, high-velocity components forming wings in the integrated H$\alpha$ profiles of (giant) \HII~regions in galaxies. These wings typically show symmetric separations from the main component between 45 and 80 km s$^{-1}$, with velocity dispersions ranging from 18 to 35 km s$^{-1}$. These high-velocity features are generally interpreted as evidence of expanding shells within the \HII~regions, which could be driven by stellar winds. However, because we resolve the \HII~region and integrate the profiles on small scales, this interpretation seems unlikely in our case.

\cite{Henney98} demonstrated (for the [OIII] 5007 \AA~line) that dust scattering of emission lines in \HII~regions can produce broad ($\sigma$ $\sim$ 13 km s$^{-1}$) redshifted components at velocities corresponding to the relative velocity between the dust and the emitting gas. In this framework, we could speculate that the $-$35~km s$^{-1}$ component is due to dust scattering. It is faint, follows dust features in Field 2, and is barely detected in Field 5. However, the expected relative velocity between the dust (assumed to be at the molecular cloud velocity of $\sim$ $-$56 km s$^{-1}$) and the emitting gas (H$\alpha$ emission at $\sim$ $-$72 km s$^{-1}$) is about 16 km s$^{-1}$, while the component is shifted by a larger value. In the Orion Nebula, velocity differences up to 30 km s$^{-1}$ have been observed for back-scattered components (\citealt{Castaneda88}, \citealt{Arthur16}), which may support this scenario.

Protostars can produce jets and outflows with velocities between 150 and 400 km s$^{-1}$, and lengths ranging from a few tens of AU to several parsecs (e.g., \citealt{Eisloffel98}), forming surrounding cavities and bow shocks (e.g., \citealt{Bally07}, \citealt{Stanke99}). Additionally, \cite{Wu04} highlighted that outflows from high-mass stars are less collimated than those from low-mass stars. In NGC 7538, \cite{rigby24} identified several molecular outflows that extend beyond the bulk velocity range (-70 to $-$40 km s$^{-1}$). \cite{Sandell09} showed that the free-free emission from IRS1 is dominated by a collimated ionized wind driving a north-south jet. \cite{Sandell20} also demonstrated that IRS1 is driving a large molecular outflow, with the blueshifted northern lobe sculpting a cavity extending to $\sim$ 280\arcsec~(3.6 pc) from IRS1. Indeed, \cite{Kraus06}, \cite{Sandell10}, and \cite{Sandell20} showed that IRS1 drives a large fan-shaped ionized and molecular (precessing) outflow, which could have cleared a cavity in NGC 7538. The western boundary of this cavity is marked by a pronounced ridge-like filament connecting IRS1-3 with IRS4 and extending toward IRS5 (corresponding to the southwest rim). The velocity range of the blueshifted emission is from $-$82 to $-$64 km s$^{-1}$, while the redshifted emission is from $-$52 to $-$31 km s$^{-1}$. In this context, the $-$92 km s$^{-1}$ component, being very local and situated within the area of our outflow model, could be attributed to an extreme velocity component associated with the outflow.

\begin{figure}[t]
\begin{center}
  \includegraphics[scale=0.4,viewport = 46 155 566 638, clip]{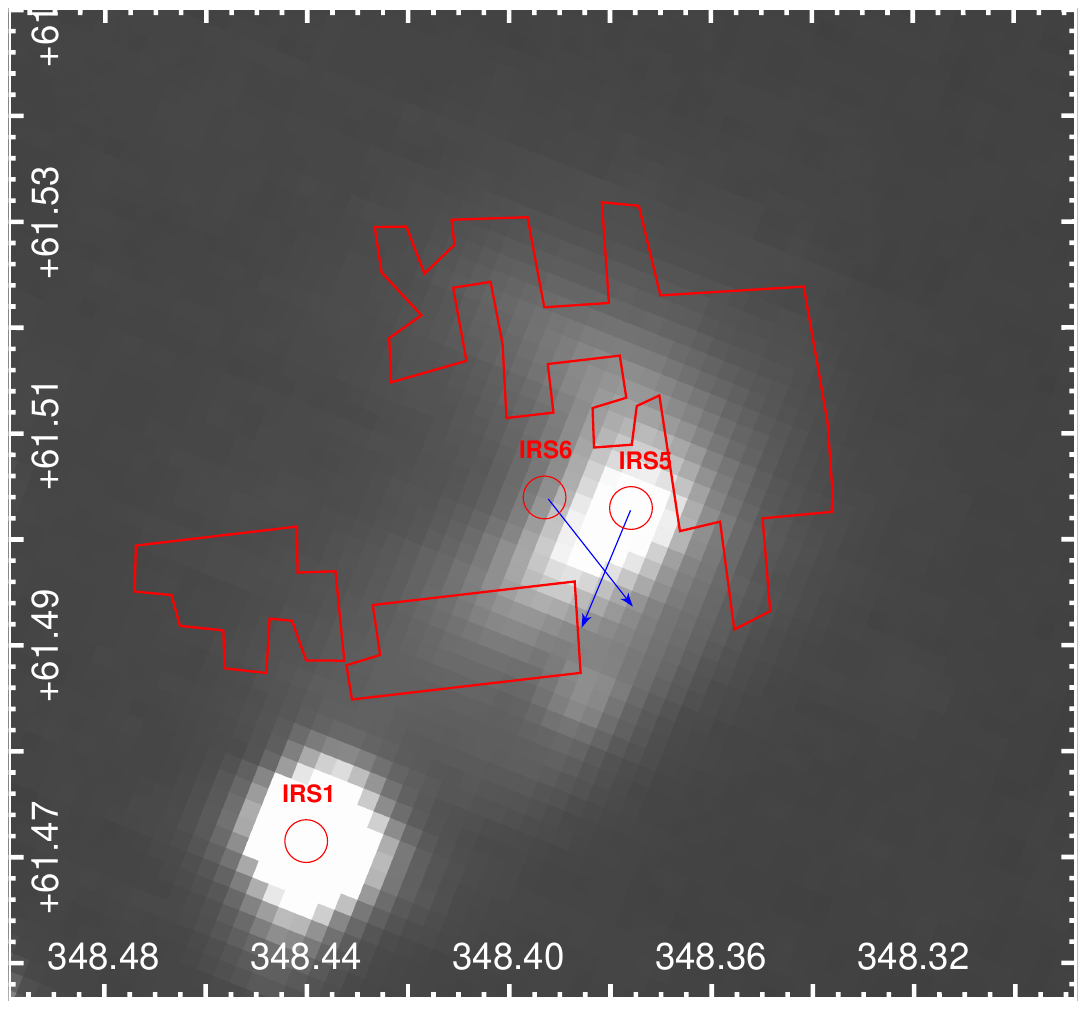}
  \caption{\label{MSXBS} MSX-21.3$\mu$m image (with decimal equatorial J2000 coordinates). The $-$35 km s$^{-1}$ component mask is overlaided (red contours). The position of IRS1, IRS5 and IRS6 are indicated by red circles. The transverse velocity direction of IRS6 and IRS5 are indicated by blue arrows (arbitrary length). }
  \end{center}
\end{figure}

The nature of the $-$35 km s$^{-1}$ component remains difficult to interpret, either as dust scattering or as linked to the outflow. An alternative explanation is that it is associated with a wind-bow shock formed around IRS6. Indeed, \cite{Mackey15} showed that such interactions can cause strong acceleration of the gas (greater than 20 km s$^{-1}$), which could explain the unusual velocity of this feature relative to the rest of the region. Furthermore, this component spatially coincides with the mid-infrared MSX 21.3 $\mu$m emission (the MIPS 24 $\mu$m and WISE 22 $\mu$m images cannot be used because they are saturated; Figure \ref{MSXBS}), which is expected to trace the extent of a stellar wind bubble (\citealt{Mackey16}). Moreover, the transverse velocity of IRS6 in the local reference frame, calculated from GAIA DR3 data and following \cite{Russeil20}, is $\sim$7 km s$^{-1}$ in the direction of the mid-infrared arc (Figure \ref{MSXBS}).

\section{Conclusion}
\label{conclusec}

The morphology of the \HII~region NGC 7538, particularly the arc-shaped structures in its northern part, and its kinematics have long been of great interest in terms of understanding the interaction between \HII~regions and their parent molecular clouds, as well as the possible triggering of stellar formation. Although the Champagne flow in the region has been known for a long time (e.g., \citealt{Deharveng79}, \citealt{Barriault07}), several additional aspects have emerged more recently: the presence of overlapping bubbles (\citealt{Beuther22}), the existence of a second PDR and an associated photon-leakage process (\citealt{Luisi16}), and a predominantly molecular, broad, precessing outflow originating from IRS1 on the southern edge of the region (e.g., \citealt{Sandell20}).

While the present study of the H$\alpha$ kinematics is consistent with previous results, since a large-scale flow is also highlighted here by the presence of a blue-shifted stream of ionised gas with a velocity exceeding 11 km s$^{-1}$, the other processes are more difficult to identify. In the central part of the region, the observed kinematics could be interpreted as the superposition of the outflow driven by IRS1 and a wind bow shock formed around IRS6 and this could be in line with identified two distinct gas flows in the inner region of NGC 7538 observed by \cite{Barriault07}. In parallel, our structure function analysis indicates that the turbulence can be shock-dominated, with a characteristic scale length between 0.72 and 1.46 pc corresponding to the typical size of bubbles observed by \cite{Beuther22}. In addition, between the inner and outer PDRs, we detect ionized gas associated with the \HII~region, consistent with findings by \cite{Luisi16}. 
Finally, we observe that profiles originating from extinction dark features or from outside the main nebula show $\sigma$ values greater than 15 km s$^{-1}$ and then larger than the typical velocity dispersion observed in the Warm Interstellar Medium (WIM). This result underlines the importance of turbulence for radiation leakage as emphasized by \cite{Kakiichi21}. 

A future objective would be to refine these results through complementary high-resolution radio continuum observations (e.g., using LOFAR data), which would help disentangle the emission contributions from the \HII~region and the outflow. In parallel, optical nebular line observations would allow us to use line ratio diagnosis diagrams (\citealt{Baldwin81}) to better constrain the ionization mechanisms of the nebula.

\begin{acknowledgements}
  We used Mistral AI (Mistral AI, 2024) and chatGPT (OpenAI, 2024) to help improve the text. HP thanks the
Laboratoire d’Astrophysique de Marseille (LAM) and Aix Marseille Université (AMU) for the financial support during his staying at LAM in December 2023. HP also thanks Cnpq for the financial support from the CNPq/MCTI/FNDCT Process N$^o$ 404160/2025-5 (44/2024 - UNIVERSAL). AZ thanks the support of the Institut Universitaire de France. We thank the referee for their comments that have helped to improve the clarity of the paper.
\end{acknowledgements}

\bibliographystyle{aa}

\end{document}